\documentclass[%
 reprint,
 amsmath,
 amssymb,
 aps,
 prd,
 nofootinbib,
 superscriptaddress
]{revtex4-2}
\makeatletter
\renewcommand\p@subsection{}      
\renewcommand\p@subsubsection{}   
\makeatother

\usepackage[normalem]{ulem}
\usepackage{graphicx}
\usepackage{xcolor}
\usepackage{mathtools}
\usepackage{tabularray}
\usepackage{multirow}
\UseTblrLibrary{booktabs}
\graphicspath{{./}{figures/}}
\usepackage{verbatim} 
\usepackage{bm}
\usepackage{hyperref}
\usepackage{orcidlink}
\hypersetup{
    colorlinks=true,
    citecolor=blue,
    urlcolor=blue,
}

\begin{document}

\title{Mixed Dark Matter: Limits from the Milky Way Satellite Galaxies
}

\author{Wendy Crumrine\orcidlink{0009-0002-7428-7019}}
\email{crumrine@usc.edu}
\affiliation{Department of Physics $\&$ Astronomy, University of Southern California, Los Angeles, California, 90007, USA}

\author{Dominic Pang}

\affiliation{Department of Physics $\&$ Astronomy, University of Southern California, Los Angeles, California, 90007, USA}

\author{Ethan O.~Nadler\orcidlink{0000-0002-1182-3825}}

\affiliation{Department of Astronomy \& Astrophysics, University of California, San Diego, La Jolla, California 92093, USA}

\author{Andrew Benson\orcidlink{0000-0001-5501-6008}}

\affiliation{Carnegie Observatories, 813 Santa Barbara Street, Pasadena, California 91101, USA}
\author{Vera Gluscevic\orcidlink{0000-0002-3589-8637}}
\email{vera.gluscevic@usc.edu}
\affiliation{Department of Physics $\&$ Astronomy, University of Southern California, Los Angeles, California, 90007, USA}

\affiliation{School of Natural Sciences, Institute for Advanced Study, Princeton, NJ 08540, USA}
\affiliation{Center for Computational Astrophysics, Flatiron Institute, New York, NY 10010, USA}

\begin{abstract}
The Standard Model of particle physics contains a diverse set of particle species, motivating the possibility of a similarly complex dark sector. Here we study two-component dark matter (DM) mixtures, in which one component behaves as standard CDM while the other suppresses the formation of small-scale structure, either through an astrophysically relevant de~Broglie wavelength (fuzzy DM; FDM) or collisional damping from temperature-independent scattering (interacting DM; IDM). Using the observed population of Milky Way satellite galaxies, we derive new leading constraints on the parameter spaces of mixed FDM and of mixed IDM coupled to photons ($\gamma$-DM), neutrinos ($\nu$-DM), or baryons ($p$-DM), for beyond-CDM fractions down to $50\%$. We require that the linear matter power spectra of allowed models remain less suppressed than a constrained reference model. The resulting $95\%$ confidence bounds on FDM mass and IDM cross section weaken systematically with decreasing fraction, following distinct power-law scalings. At $50\%$ fraction, IDM cross section bounds weaken by a factor of $\sim$2--6 and FDM mass bounds by $\sim$1.5, relative to the $100\%$ case. We forecast that idealized future satellite surveys, which adopt approximate LSST sensitivity thresholds, can improve these $100\%$ bounds by a factor of $\sim$1.6--14 for IDM and $\sim$3 for FDM. Self-consistent cosmological simulations of mixed DM scenarios will be essential to more robustly characterize the degeneracy between particle physics parameters and fractional contribution, to extend constraints to lower fractions, and to identify signatures beyond satellite abundance to further inform these models.

\keywords{Cosmology - Dark matter - Particle dark matter - Particle astrophysics - Milky Way}
\end{abstract}

\maketitle

\section{Introduction}\label{sec:intro}

The standard cosmological paradigm assumes that dark matter (DM) is cold, collisionless, and interacts solely through gravity. This cold dark matter (CDM) framework broadly reproduces the large-scale structure of the Universe, as probed by observations of the cosmic microwave background (CMB), galaxy clustering, and the Lyman-$\alpha$ forest \cite{2020,Chabanier_2019}---though notable tensions leave room for possible beyond-CDM extensions~\cite{Abdalla_2022, Di_Valentino_2021, Perivolaropoulos_2022}. At small, nonlinear scales, recent studies also find consistency between CDM predictions and observed halo abundances down to $\sim$$10^8$ $M_\odot$, as inferred from the observed population of Milky Way (MW) satellite galaxies \cite{Nadler_2020,Jethwa_2017}. Upcoming and ongoing observations aim to provide additional, independent tests of small scale clustering, using strong gravitational lensing \cite{Hsueh_2019, Gilman_2019,Keeley:2024brx}, stellar streams \cite{Banik_2018, Bonaca_2019}, and dwarf galaxy stellar velocity dispersions \cite{Kim_2021,Esteban_2023}.

Current observational and theoretical uncertainties below halo masses of $\sim$$10^8$ $M_\odot$ leave room for beyond-CDM cosmologies, in which small-scale structure can differ substantially from CDM. This flexibility reflects both observational incompleteness and limited understanding of the baryonic processes that influence dwarf-galaxy formation and evolution, including tidal and ram-pressure stripping and feedback from supernovae, stellar radiation, and reionization \cite{bullock2010notesmissingsatellitesproblem, Bullock_2017}.

Many such beyond-CDM scenarios modify small-scale matter density fluctuations by suppressing the formation of halos, and the dwarf galaxies they host, below a characteristic mass scale. Commonly proposed microphysical mechanisms include DM free streaming out of gravitational potential wells (in the case of warm dark matter, WDM; \cite{1988ApJ...332....1S,Sommer_Larsen_2001}), an astrophysically relevant DM de Broglie wavelength (fuzzy dark matter, FDM; \cite{Hu_2000}), and the collisional damping and acoustic oscillations that arise from momentum exchange between DM and Standard Model (SM) particles, radiation, or dark radiation (interacting dark matter, IDM; \cite{Boehm_2005,Dvorkin_2014,Vogelsberger_2016,Boddy_2018,Nadler_2019_b,Maamari_2021,nadler2024cozmicicosmologicalzoomin,An_2025, zhou2024searchingdarkmatterinteractions}). 

In this analysis, we focus on FDM and on IDM that elastically scatters with SM photons, neutrinos, or protons in the early Universe. We extend previous work by allowing only a fraction ($f_\varphi$ for FDM, $f_\chi$ for IDM) of the total DM to exhibit non-standard properties, while the remainder behaves as CDM. We derive observational constraints on all four scenarios across a range of these fractions, entering a region of parameter space largely unexplored. With some exceptions~\cite{Rogers_2023,Rogers_2025, winch2025highredshiftsmallscaletestsultralight,lazare2025constraintsfuzzydarkmatter, stuart2025constraintsdarkmatterbaryoninteraction, Zu_2026, Hooper_2022_Likelihood, He_2023, Stadler_2020, straight2026cmbconstraintsdarkmatterproton}, previous analyses have focused primarily on scenarios with $f_\varphi =1, \,f_\chi = 1$, employing a wide variety of observables. For FDM, constraints derive from MW satellite abundances~\cite{nadler2024cozmicicosmologicalzoomin,nadler2026warmfuzzygeneralizedultralight}, Lyman-$\alpha$ forest flux power spectra~\cite{Rogers_2021}, stellar heating~\cite{dalal2022fuzzyexcludingfdmsizes}, and dwarf galaxy kinematics~\cite{Zimmermann_2025}. For IDM, bounds come from CMB anisotropies~\cite{Wilkinson_2014_nu,Wilkinson_2014_photon,Escudero_2015,Stadler_2018,Paul_2021}, Lyman-$\alpha$~\cite{Hooper_2022,He_2023,He_2025,Rogers_2022, Hooper_2022_Likelihood}, MW satellites~\cite{B_hm_2001,B_hm_2014,Escudero_2018,Akita_2023,Crumrine_2025,nadler2024cozmicicosmologicalzoomin}, cosmic reionization~\cite{Dey_2023}, high-energy astrophysical probes~\cite{Ferrer_2023,Fujiwara_2024,Cline_2023,Cline_2023_b}, galaxy cluster thermodynamics~\cite{stuart2025constraintsdarkmatterbaryoninteraction}, dwarf-galaxy density profiles~\cite{heston2024constraining}, and combined cosmological analyses~\cite{Becker_2021,Brax_2023,Brax_2023_1,giare2023hints}.

For our FDM analysis, we assume the beyond-CDM component is an ultralight scalar field whose wave dynamics are controlled by the particle mass, $m_\varphi$. For IDM, we assume a temperature-independent momentum-transfer cross section, $\sigma_{\chi j}$, where $j\in{\{\gamma,\,\nu,\,p\}}$ denotes the scattering target---photon, neutrino, or proton, respectively. This cross-section parametrization is model-independent, but concrete realizations include: a millicharged DM component that scatters elastically with photons in a Thomson-like manner, couplings between DM and sterile neutrinos \cite{Brax_2023_1,Brax_2023,Paul_2021}, or DM that scatters elastically with protons through a contact interaction via a heavy mediator \cite{Gluscevic_2018, Boddy_2018_EFT}.

In all four scenarios (FDM, $\gamma$-DM, $\nu$-DM and $p$-DM) and for the parameter ranges and scales of interest, the key physics driving departures from CDM operates before non-linear structures form in the universe; the effect of each scenario is thus a change in the initial conditions for structure formation in the form of a scale-dependent alteration of the linear matter power spectrum, $P(k)$. We therefore map existing observational constraints on $P(k)$ onto the four scenarios to arrive at bounds on interaction strength and mass of each DM subcomponent. Specifically, we use constraints derived using the Cosmological ZooM-in simulations with beyond-CDM Initial Conditions II (COZMIC II) suite \cite{An_2025}. The COZMIC II suite consists of DM-only MW-like zoom-in simulations with suppressed $P(k)$ that emulate mixed DM scenarios. Its analysis provides constraints on the onset of the linear $P(k)$ suppression and its amplitude at large wavenumber $k$, using the observed abundance of MW satellites from the Dark Energy Survey (DES) and Pan-STARRS1 (PS1)~\cite{Drlica_Wagner_2020}.

Applying new approaches to mapping COZMIC II bounds onto the mixed DM scenarios considered here, we derive new constraints on $m_\mathrm{\varphi}$ for FDM and on $\sigma_{\chi j}$ for all three IDM scenarios, for fractional contributions $f_\varphi$ and $f_\chi$ down to $50\%$. These limits exclude models that significantly suppress power at scales corresponding to the halos that host the faintest observed dwarf galaxies; i.e., halos of order $\gtrsim
10^8$$M_{\mathrm{\odot}}$, and wavenumbers $k$ $\lesssim$ $10$ $h~\mathrm{Mpc}^{-1}$ in linear theory (see Eq.\ (5) from Ref.~\cite{Nadler_2019_b}). Our results show constraints weakening systematically
with decreasing $f_\varphi$ and $f_\chi$, following
power-law scalings that reflect degeneracies between the
beyond-CDM fraction and the model parameters ($m_\varphi$
for FDM, $\sigma_{\chi j}$ for IDM, the latter at fixed $m_\chi$).

This paper is organized as follows: In Section \ref{sec:theory}, we review our theoretical framework for modeling FDM, $\gamma$-DM, $\nu$-DM and p-DM, extending to scenarios in which only a fraction of DM exhibits non-standard properties. In Section \ref{sec:approach}, we present our methods for deriving constraints on FDM mass for a range of interaction fractions, and on interaction strength ($\gamma$-DM, $\nu$-DM and $p$-DM) for a range of DM masses and fractions. Section~\ref{sec:results} summarizes our results and compares them with idealized future-satellite-survey forecasts, followed by discussion and conclusions in Section~\ref{sec:conclusion}. 

Throughout, we adopt the following cosmological parameters, following Refs.~\cite{Maamari_2021,Crumrine_2025}: Hubble constant $h=0.6932$, baryon density $\Omega_bh^2=0.02223$, DM density $\Omega_{\mathrm{dm}}h^2=0.1153$, optical depth to reionization $\tau_{\mathrm{reio}}=0.081$, scalar perturbations amplitude $A_s=2.464\times10^{-9}$, scalar spectral index $n_s=0.9608$, and effective number of neutrino species $N_{\mathrm{eff}}=3.046$. We assume massless neutrinos when deriving our constraints, following Ref.~\cite{Crumrine_2025}.

\section{Theory}\label{sec:theory}

In this section, we review the theoretical foundations of FDM and IDM, focusing on the mechanisms by which each suppresses small-scale structure and the linear matter power spectrum. We extend both frameworks to mixed scenarios, treating each DM subcomponent as a separate fluid within the Einstein-Boltzmann system. We begin with FDM in Section~\ref{sec:fdm} and then explore all three IDM scenarios ($\gamma$-DM, $\nu$-DM, and $p$-DM) in Section~\ref{sec:idm}.

\subsection{Fuzzy Dark Matter}\label{sec:fdm}
The FDM scenario treats DM as an ultralight scalar field, $\varphi$, with a mass $m_\varphi\,$$\sim$$\,10^{-22}\,\mathrm{eV}$, and whose evolution is governed by the Klein--Gordon equation \cite{Passaglia_2022},
\begin{equation}
    \varphi^{\prime\prime} + 2\mathcal{H}\varphi^\prime + a^2 m_\varphi^2 \varphi = 0,
\end{equation} where $a$ is the scale factor, $\mathcal{H}$ is the conformal Hubble rate, and primes denote derivatives with respect to conformal time. When $m_\varphi \gg \mathcal{H}/a$ (the physical Hubble rate), $\varphi$ oscillates, and its time-averaged stress-energy behaves effectively as a fluid. To model linear cosmological perturbations in this regime, we therefore replace the standard Boltzmann equations with the following fluid equations~\cite{Ma_1995,Marsh_2016}:
\begin{equation}\label{boltzmann_fdm}
\begin{aligned}
\delta_\varphi' &= -\theta_\varphi + 3\phi' - 3\mathcal{H}c_{\mathrm{eff}}^2\,\delta_\varphi, \,\\
\theta_\varphi' &= -\mathcal{H}\theta_\varphi + c_{\mathrm{eff}}^2\, k^2 \delta_\varphi + k^2\psi \,.
\end{aligned}
\end{equation}
Here $\delta_\varphi$ is the scalar field density perturbation, $\theta_\varphi$ is its velocity divergence, $\psi$ and $\phi$ are scalar metric perturbations, $k$ is the comoving wave number, and $c_{\mathrm{eff}}^2$ is the sound speed of the effective fluid, given by
\begin{equation}\label{c_eff}
    c_{\mathrm{eff}}^2 = \frac{k^2}{4a^2 m_\varphi^2 + k^2}.
\end{equation}
For large $m_\varphi$, $c_{\mathrm{eff}}^2 \to 0$ and perturbations grow as in standard CDM at all scales of interest. However, for ultralight $m_\varphi$, the field's de~Broglie wavelength can reach astrophysical scales, inducing an effective ``quantum" pressure (encoded in $c_{\mathrm{eff}}^2$) that suppresses the growth of perturbations \cite{Hu_2000,Marsh_2016}. Specifically, on small scales ($k \gg 2am_\varphi$), quantum pressure dominates ($c_{\mathrm{eff}}^2 \to 1$), suppressing the growth of fluctuations. However, on large scales ($k \ll 2am_\varphi$), $c_{\mathrm{eff}}^2$ vanishes and perturbations grow as in CDM.  

This scale-dependent suppression defines a characteristic Jeans wavenumber \cite{Hu_2000,Passaglia_2022},
\begin{equation}\label{jeans_fdm}
    k_J \sim \sqrt{am_\varphi \mathcal{H}}\,,
\end{equation}
above which ($k > k_J$) structure formation is inhibited. Values of $m_\varphi$ that place $k_J$ at scales relevant for MW satellites can be ruled out, as they would prevent the growth of structure at observed scales.

A mixed scenario is specified by $f_\varphi \in [0,1]$, the FDM share of the total background DM density:
\begin{equation}\label{fraction_split_fdm}
\rho_{\varphi} = f_\varphi\,\rho_{\rm dm}, \qquad \rho_{\rm CDM} = (1 - f_\varphi)\,\rho_{\rm dm}.
\end{equation}
Within the perturbation framework, the two subcomponents evolve according to separate fluid equations: FDM follows the full Eq.~(\ref{boltzmann_fdm}), while CDM follows the same equation without the pressure terms, as expected for a cold, collisionless fluid. The total DM perturbation is then
\begin{equation}\label{Delta_Tot_FDM}
\delta_{\rm dm} = (1 - f_\varphi)\,\delta_{\rm CDM} + f_\varphi\,\delta_{\varphi},
\end{equation}
which enters observables quadratically through $P(k) \propto \langle|\delta_m(k)|^2\rangle$, where $\delta_m$ includes contributions from both DM and baryons.

Quantum pressure damps $\delta_\varphi$ relative to $\delta_{\rm CDM}$ on small scales, so reducing $f_\varphi$ shifts weight toward the unsuppressed CDM component and mitigates the small-scale departure from CDM in $\delta_{\rm dm}$. Note that $f_\varphi$ also couples (via $\delta_\varphi$ and $\theta_\varphi$) to the total stress-energy that sources the metric perturbations $\psi$ and $\phi$ through the gravitational Poisson constraint; this feedback further shapes the evolution of $\delta_{\rm dm}$.

\subsection{Interacting Dark Matter (IDM)}\label{sec:idm}

To model early Universe linear perturbations in an IDM cosmology, collisional terms are introduced throughout the standard Boltzmann equations, to capture the induced drag forces and pressure support. For a detailed treatment  in this regime, see Refs.~\cite{Dvorkin_2014,Becker_2021}. For $p$-DM, the relevant equations are, 
with boxed terms indicating scattering-induced departures from standard CDM, \begin{equation}\label{boltzmann_baryon_DM_final}
\begin{aligned}
\delta_{b}^\prime &= -\theta_{b} + 3\phi^\prime, \\[3pt]
\theta_{b}^\prime &= -\mathcal{H}\theta_{b} + c_{b}^{2}k^{2}\delta_{b} + k^2\psi \\[-2pt]
                  &\quad +\,\Gamma_{b\gamma}(\theta_{\gamma}-\theta_{b})
                         -\,\boxed{\Gamma_{b\chi}(\theta_{b}-\theta_{\chi})}, \\[6pt]
\delta_{\chi}^\prime &= -\theta_{\chi} + 3\phi^\prime, \\[3pt]
\theta_{\chi}^\prime &= -\mathcal{H}\theta_{\chi} 
                      + \boxed{c_{\chi}^{2}k^{2}\delta_{\chi}}
                      + k^2\psi 
                      -\,\boxed{\Gamma_{\chi b}(\theta_{\chi}-\theta_{b})}.
\end{aligned}
\end{equation} $\delta_b$ is the baryon density perturbation, treated here as proton-dominated, while $\delta_\chi$ is its DM counterpart. 
$\theta_b$ ($\theta_\chi$, $\theta_\gamma$) denotes the baryon (DM, photon) velocity divergence, 
and $c_b$ ($c_\chi$) is the sound speed of the baryon (DM) fluid. 

The rate at which scattering transfers momentum to baryons is given by $\Gamma_{b \gamma}$ and $\Gamma_{b \chi}$, and to DM by $\Gamma_{\chi b}$. The first represents the standard photon-baryon coupling due to Thomson scattering, while the latter two arise from IDM interactions and are related by momentum conservation; i.e., $\Gamma_{\chi b} = (\rho_b / \rho_\chi)\,\Gamma_{b \chi}$, giving \cite{Dvorkin_2014}
\begin{align}\label{momentum_proton}
    \Gamma_{\chi b} &= a\,\frac{\rho_b\,\sigma_{\chi b}\,\mathcal{N}_0\,Y_{\rm b}}{m_\chi + m_b}
     \!\left[\frac{T_b}{m_b} + \frac{T_\chi}{m_\chi} + \frac{V_{\chi b}^2}{3}\right]^{1/2}.
\end{align} Here, $\rho_b$ ($\rho_\chi$) is the baryon (DM) energy density, $\sigma_{\chi b}$ denotes the momentum-transfer cross section between DM and baryons, $m_\chi$ and $m_b$ are the DM and baryon particle (proton) masses, $T_\chi$ and $T_b$ are their corresponding temperatures, and $V_{\chi b}$ is the bulk relative velocity between the DM and baryon fluids. The coefficient $\mathcal{N}_0 = 8\sqrt{2}/(3\sqrt{\pi}) \approx 2.13$ arises from thermal averaging over the Maxwell-Boltzmann distribution, and $Y_{\rm b}$ accounts for the proton (hydrogen) fraction.

For $\gamma$–DM, the Boltzmann equation modifications are analogous to the proton case and are given by \begin{equation}\label{boltzmann_rad}
\begin{split}
    \delta_{\gamma}^\prime &= -\frac{4}{3}\theta_{\gamma} + 4\phi^\prime, \\
    \theta_{\gamma}^\prime &= k^2\!\left(\frac{1}{4}\delta_\gamma - \sigma_\gamma\right) + k^2\psi
    - \Gamma_{\gamma b}\!\left(\theta_{\gamma}-\theta_{b}\right)^{\dagger}
    - \boxed{\Gamma_{\gamma\chi}\!\left(\theta_{\gamma}-\theta_{\chi}\right)}, \\[3pt]
    \delta_{\chi}^\prime &= -\theta_{\chi} + 3\phi^\prime, \\
    \theta_\chi^\prime &= -\mathcal{H}\theta_{\chi}
    + \boxed{c^{2}_{\chi}k^{2}\delta_{\chi}}
    + k^2\psi
    - \boxed{\Gamma_{\chi \gamma}\!\left(\theta_{\chi}-\theta_{\gamma}\right)}.
\end{split}
\end{equation} Here the beyond-CDM terms are again boxed, $\sigma_\gamma$ is the shear stress potential, and momentum transfer---between DM and the photon fluid---is given by

\begin{equation}\label{momentum_rad}
    \begin{split}
        \Gamma_{\chi \gamma} = \frac{4\rho_\gamma}{3\rho_\chi}\Gamma_{\gamma \chi} = \frac{4\rho_\gamma}{3\rho_\chi}n_\chi\sigma_{\chi \gamma} ac = \frac{4\rho_\gamma}{3m_\chi}\sigma_{\chi \gamma}ac,
    \end{split}
\end{equation} 
where $\rho_\gamma$ is the photon energy density, $n_\chi$ is the DM number density, and $c$ is the speed of light. The $\nu$-DM equations follow identically, excluding the daggered term, which corresponds to the standard photon-baryon coupling and arises only in the photon case. 

While Eqs.~\ref{momentum_proton} and \ref{momentum_rad} show the momentum-exchange rate's explicit dependence on $\sigma_{\chi j}$ and $m_\chi$ (here with $j\in{\{\gamma, \nu, p\}}$), IDM literature often expresses this dependence through a single dimensionless parameter, given by
\begin{equation}\label{u}
    \begin{split}
    u_{\chi j}=\frac{\sigma_{\chi j}}{\sigma_{T}}\left(\frac{m_\chi}{100~\mathrm{GeV}}\right)^{-1},
    \end{split}
\end{equation} 
where $\sigma_{T}$ is the Thomson scattering cross section (i.e., $6.65\times10^{-25}$ $\textrm{cm}^2$). The cross section itself can be parametrized as $\sigma_{\chi j} = \sigma_0 T_\chi^n$ 
for radiation scattering ($\sigma_{\chi b} = \sigma_0 v^n$ for baryon scattering), with $n \in \{-4, -2, 0, 2, 4, 6\}$ encompassing a variety of microphysical models \cite{B_hm_2001,Boddy_2018_EFT,Becker_2021}. Here we consider only the temperature-independent (velocity-independent) case $n=0$, deferring other values to future work.

Although the three IDM scenarios under investigation differ in target, suppression of density perturbations arises from the same two underlying mechanisms---collisional damping and acoustic pressure support. For high $m_\chi$, the former is the leading effect---arising from momentum exchange that couples DM velocity to the scattering target, and captured by the boxed terms, $\Gamma_{\chi j}(\theta_\chi-\theta_j)$, in Eqs.~\ref{boltzmann_baryon_DM_final} and \ref{boltzmann_rad}. At low $m_\chi$, pressure support dominates---arising from a scattering-induced non-negligible sound speed, $c_\chi$, in the DM fluid, and governed by the boxed pressure term, $c_\chi^2 k^2\delta_\chi$, in Eqs.~\ref{boltzmann_baryon_DM_final} and \ref{boltzmann_rad}. Here $c_\chi$ is given by 
\begin{equation}\label{sound speed}
    \begin{split}
    c^2_\chi=\frac{k_BT_{\chi}}{m_\chi}\left(1-\frac{1}{3}\frac{\partial \log T_{\chi}}{\partial \log a}\right),
    \end{split}
\end{equation} 
where $k_B$ is Boltzmann's constant and $T_\chi$, the DM fluid temperature, is driven toward that of the warmer scattering target, $T_j$, via
\begin{equation}\label{Temp}
\begin{split}
T_\chi^\prime=-2 \mathcal{H} T_{\chi} - 2 \Gamma_{\chi j} \left(T_{\chi} - T_{j} \right).
\end{split}
\end{equation} We adopt a non-relativistic treatment of $c_\chi$, appropriate since the physics relevant to MW satellite-scale suppression (i.e., at redshifts $\sim$$10^6$ and later) occurs when $c_\chi$ is below $\sim$$0.1c$. For $\gamma$-DM and $\nu$-DM, we follow previous literature and neglect alterations in temperature of the radiation arising from scattering, modeling its evolution as in the standard CDM cosmology, where $T_j=T_{j,0}(1+z)$. Specifically, $T_{\gamma,0}=2.73$K (the CMB temperature today) while $T_{\nu,0}$ becomes $1.95$K when accounting for the usual factor of $(4/11)^\frac{1}{3}$.

Depending on which suppression mechanism dominates for a given model, observational consistency requires a different epoch of kinetic decoupling (i.e., the era in which momentum exchange between DM and its target becomes negligible). For high $m_\chi$ along our bound---where collisional damping dominates---the scale of linear suppression is set by the causal horizon size at decoupling. Since perturbations corresponding to the faintest MW satellite hosts ($\sim$$10^8 M_
\odot$ subhalos) enter the horizon around redshift $z\sim10^6$, decoupling must occur at or before this to avoid impacting the observed satellite population.\footnote{To obtain this redshift estimate, we map $10^8$ $M_\odot$ to its comoving length scale (see Eq.\ (5) of Ref.~\cite{Nadler_2019_b}), then set this equal to the horizon size, solving for redshift. For high $m_\chi$ models, this decoupling redshift is also recovered from Eqs.~\ref{momentum_proton} and \ref{momentum_rad} by finding when the momentum-transfer rate falls below the Hubble rate.} However, at low $m_\chi$---where acoustic pressure dominates---the scale of linear suppression is instead set by the Jeans length of the DM fluid, given by \cite{Kolb:1990vq}
\begin{equation}
\lambda_J \simeq c_\chi \left(\frac{\pi}{G\rho_{\chi}}\right)^{1/2},
\end{equation}
where $G$ is Newton's gravitational constant. Here, $\lambda_J$ grows with $c_\chi$, tracking $T_\chi$ via Eq.~(\ref{sound speed}). $T_\chi$ remains elevated during coupling and persists above adiabatic expectations even afterward, keeping $\lambda_J$ above MW satellite scales. Consistency with observation therefore demands earlier decoupling (around $z\sim10^9$), leaving sufficient time for $\lambda_J$ to fall below MW satellite scales prior to nonlinear collapse.

\begin{figure*}[htbp]\label{panel_nine}
    \centering
    \includegraphics[width=\textwidth]{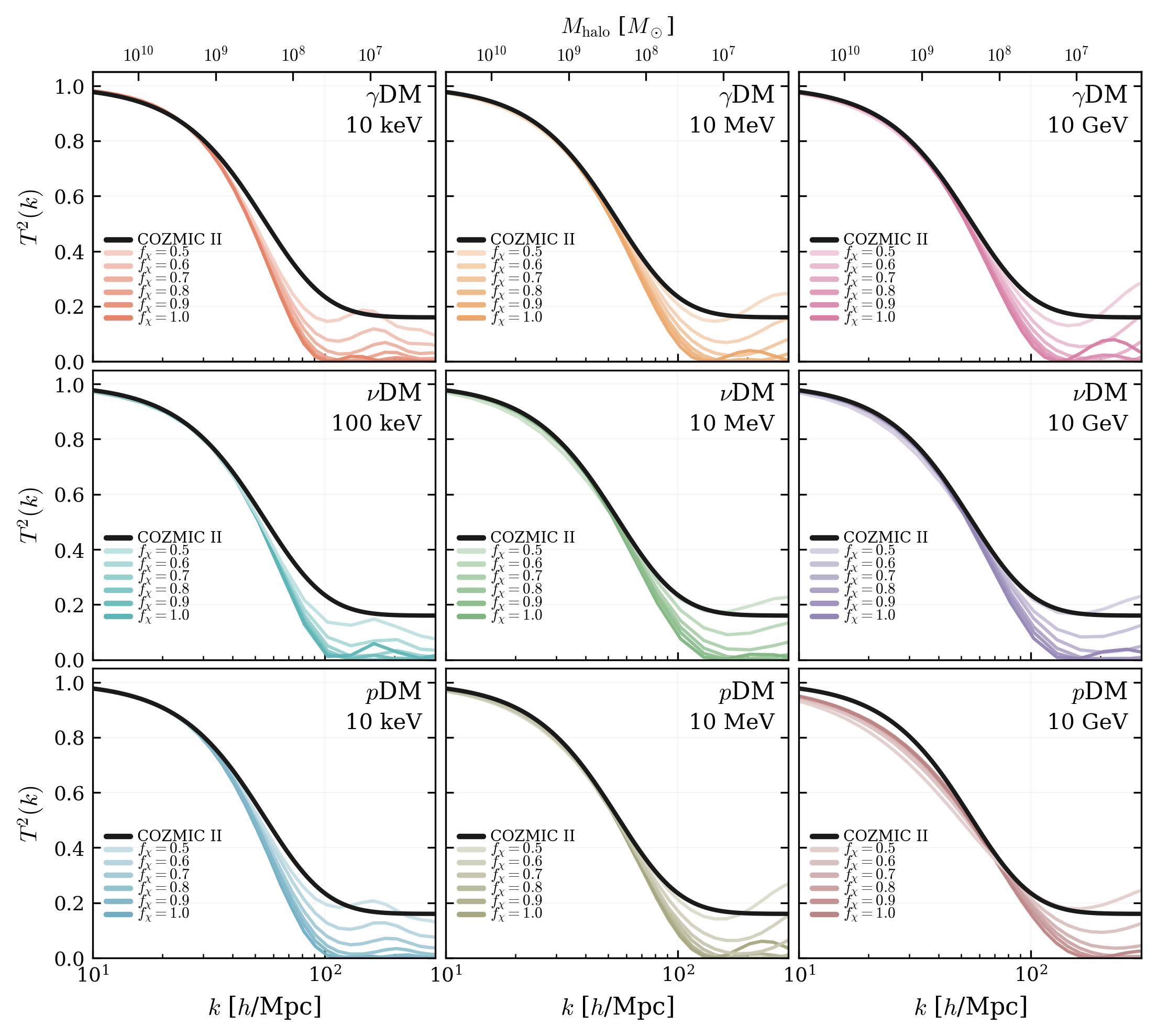}
    \caption{Transfer functions $T^2(k)$ for mixed IDM scenarios. Rows show $\gamma$-DM (top), $\nu$-DM (middle), and $p$-DM (bottom); columns span representative DM masses, illustrating how suppression morphology evolves from pressure-dominated to drag-dominated regimes. The black curve shows the COZMIC~II benchmark envelope with $\{k_\mathrm{hm},\,\delta\} = \{57\,h\,\mathrm{Mpc}^{-1},\,0.4\}$, representing the shallowest $T^2(k)$ suppression ruled out by MW satellite observations~\cite{An_2025}. Colored curves show the $95\%$ marginally excluded $T^2(k)$ from the \textit{envelope} method (Section~\ref{sec:approach}) for beyond-CDM fractions $f_\chi = 0.5$--$1.0$; models that fall below the COZMIC~II envelope at all $k$ are ruled out. The top axis shows halo masses corresponding to each wavenumber in linear theory (in units of solar masses).}
    \label{fig:Panel_9}
\end{figure*}

The IDM formalism thus far described assumes the entire DM population undergoes scattering with a target species. As in Section~\ref{sec:fdm}, we now generalize to scenarios where only a fraction of DM interacts, while the remainder behaves as standard CDM. The mixed scenario is again specified by $f_\chi \in [0,1]$, with the total DM energy density splitting as
\begin{equation}
\rho_{\rm dm} = \rho_{\rm CDM} + \rho_\chi = (1-f_\chi)\,\rho_{\rm dm} + f_\chi\,\rho_{\rm dm}.
\end{equation}
The two DM fluids evolve self-consistently within the Boltzmann equation system: CDM follows the standard equations without drag or pressure terms, while IDM evolves according to Eqs.~\ref{boltzmann_baryon_DM_final} or \ref{boltzmann_rad}.

As with FDM, $f_\chi$ governs observables primarily through the density-weighted total DM perturbation,
\begin{equation}\label{Delta_Tot_IDM}
\delta_{\rm dm} = (1-f_\chi)\,\delta_{\rm CDM} + f_\chi\,\delta_\chi,
\end{equation}
so that reducing $f_\chi$ mitigates overall small-scale suppression. However, unlike FDM, IDM introduces an indirect coupling through the target species. While the drag coefficients in Eqs.~\ref{momentum_proton} and \ref{momentum_rad} are independent of $\rho_\chi$ due to momentum conservation, the back-reaction on the target species retains its dependence on $f_\chi$. This modifies the target velocity $\theta_j$, which feeds back into IDM dynamics through the drag term $\Gamma_{\chi j}(\theta_\chi - \theta_j)$. Finally, as in FDM, both $\delta_\chi$ and $\theta_\chi$ contribute to the stress-energy tensor, coupling $f_\chi$ to the metric perturbations $\psi$ and $\phi$.

\section{Methods}
\label{sec:approach}

\begin{figure*}[htbp]\label{panel_fdm}
    \centering
    \includegraphics[width=\textwidth]{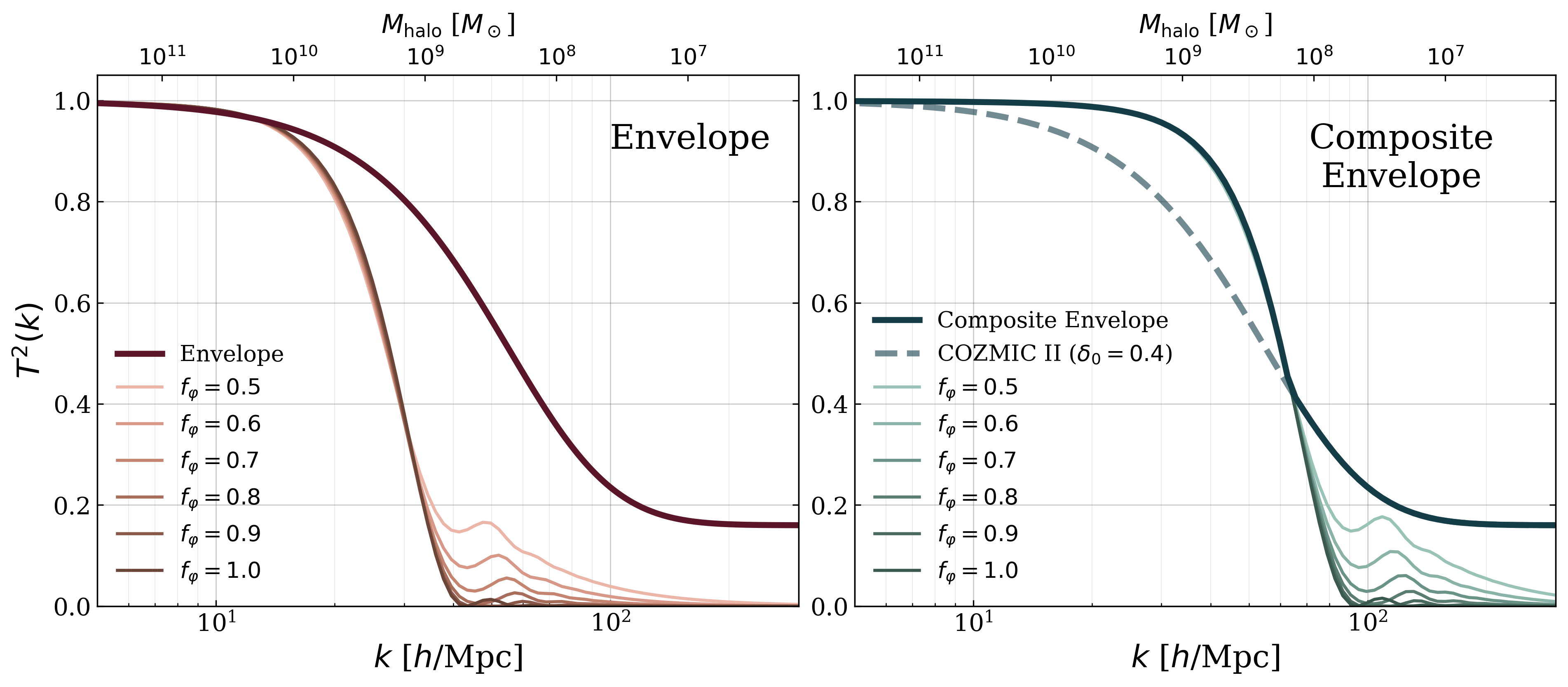}
    \caption{\textit{Left:} \textbf{Envelope Method.} FDM transfer functions $T^2(k)$ that correspond to the 95\% confidence-level lower bound on FDM particle mass are plotted for the indicated fraction of FDM, $f_\varphi$; models exhibiting greater suppression are ruled out. The reference COZMIC~II envelope ($\delta = 0.4$, $k_{\mathrm{hm}} = 57\,h\,\mathrm{Mpc}^{-1}$) from Ref.~\cite{An_2025}, representing the shallowest suppression (largest $\delta$) ruled out by MW satellite observations, is also shown (bold maroon). Each plotted FDM curve saturates the envelope criterion of Eq.~(\ref{envelope_criterion}) at the 95\% bound. \textit{Right:} \textbf{Composite Envelope Method.} This approach combines the sharp cutoff from the $f_\varphi = 1$ COZMIC~I constraint~\cite{nadler2024cozmicicosmologicalzoomin} with the high-$k$ plateau of the COZMIC~II $\delta = 0.4$ benchmark to form the composite envelope (bold blue), better capturing FDM's suppression morphology. The dashed curve (partly overplotted by the envelope) shows the original COZMIC~II benchmark envelope for reference. Models satisfying Eq.~(\ref{envelope_criterion}) with $T^2_{\mathrm{env}}$ replaced by the composite envelope are ruled out. The top axis shows halo masses corresponding to each wavenumber in linear theory (in units of solar masses).}
    \label{fig:Panel_FDM}
\end{figure*}
In all four scenarios under investigation, the key beyond-CDM physics modifies linear $P(k)$ and thus affects only the initial conditions for structure formation; non-linear growth proceeds as in CDM cosmology.\footnote{This is expected to be a good approximation in the parameter regime allowed by current CMB constraints on these scenarios~\cite{nadler2024cozmicicosmologicalzoomin, May_2023, Maamari_2021}, and with greater margin at the tighter values probed here.} We exploit this to bypass dedicated simulations and map our bounds directly from existing MW satellite constraints on $P(k)$. Specifically, we assume that any scenario whose transfer function $T^2(k) \equiv P(k)/P_{\mathrm{CDM}}(k)$ is \textit{more} suppressed than the transfer functions corresponding to the current bounds on $P(k)$ suppression is itself inconsistent with observations. 

As a benchmark for allowed $P(k)$ suppression, we rely on an analytic fractional non-CDM (f-NCDM) parameterization of $T^2(k)$ that Ref.~\cite{An_2025} uses to set the initial conditions for their COZMIC~II simulation suite of DM-only MW zoom-in simulations of MW analogs. Their analysis then forward-models the MW satellite population with an empirical galaxy-halo connection and compares the result against DES and PS1 observations via a Poisson likelihood that accounts for survey selection effects and observational incompleteness~\cite{Mao_2015,Drlica_Wagner_2020}. The analytic transfer functions used for their initial conditions take the form  
\begin{equation}\label{transfer_function}
\begin{split}
T_{f\mathrm{-NCDM}}^2&(k,\,\delta,\,k_{\mathrm{hm}}) 
= \frac{P_{f\mathrm{-NCDM}}(k,\,\delta,\,k_{\mathrm{hm}})}{P_{\mathrm{CDM}}(k)} \\
=\,\,&\bigl[(1-\delta)\,T_{\mathrm{WDM}}\!\bigl(k,k_{\mathrm{hm}}(m_{\mathrm{WDM}})\bigr)
+ \delta\bigr]^2,
\end{split}
\end{equation} where $T_{\mathrm{WDM}}\bigl(k,k_{\mathrm{hm}}(m_{\mathrm{WDM}})\bigr)$ follows the analytic thermal-relic WDM form introduced by Ref.~\cite{Viel_2005} (refined by Ref.~\cite{vogel2023enteringerameasuringsubgalactic}), which captures the characteristic suppression of linear power below the half-mode scale, $k_{\mathrm{hm}}$, defined by $T^2(k_{\mathrm{hm}})=0.25$ and set by the thermal-relic mass $m_\mathrm{WDM}$. Following Ref.~\cite{Hooper_2022_Likelihood}, Ref.~\cite{An_2025} extends $T_{\mathrm{WDM}}\bigl(k,k_{\mathrm{hm}}(m_{\mathrm{WDM}})\bigr)$ with a constant offset, $\delta$, where the value of $\delta$ increases to emulate larger CDM fractions in the mix of cosmological fluids~\cite{Boyarsky_2009}, since the cold component alone clusters on subhorizon scales prior to decoupling.

The COZMIC~II analysis sets a $95\%$ confidence-level lower bound of $k_{\mathrm{hm}} = 57~h\,\mathrm{Mpc}^{-1}$ at $\delta = 0.4$, with models at higher $\delta$ statistically indistinguishable from CDM. We adopt their $\delta = 0.4$ benchmark to probe FDM and IDM parameter spaces via two approaches: an \textit{envelope} method applicable to all scenarios, and a \textit{composite envelope} method tailored to FDM.

The \textit{envelope method} classifies models as ruled out when their transfer functions satisfy
\begin{equation}\label{envelope_criterion}
T^2(k) \leq T^2_{\mathrm{env}}(k), \quad \text{for all } k \leq k_{\max},
\end{equation}
where $T^2_{\mathrm{env}}$ corresponds to $(\delta,\,k_{\mathrm{hm}}) = (0.4,\,57\,h\,\mathrm{Mpc}^{-1})$. The 95\% confidence bound in beyond-CDM parameter space is set by the least suppressed model satisfying this criterion. Following Ref.~\cite{Maamari_2021}, we adopt $k_{\max} = 125\,h\,\mathrm{Mpc}^{-1}$, well beyond the scales where MW satellite abundance is most sensitive to $P(k)$ suppression~\cite{Nadler_2025}. Beyond $k_{\max}$, mode-crossing contributions to MW-satellite scales are assumed negligible.

While the envelope method is well suited to IDM, whose transfer functions roll off gradually through the half-mode scale and closely trace the COZMIC~II benchmark (see Fig.~\ref{fig:Panel_9}), FDM suppresses power far more steeply. The envelope method consequently yields overly conservative FDM bounds, as FDM transfer functions fail to satisfy the condition given in Eq.~(\ref{envelope_criterion}) even at small $m_\varphi$, where MW-satellite-scale power is already heavily suppressed (see Fig.~\ref{fig:Panel_FDM}, left panel).

To construct a more realistic bound, we introduce a \textit{composite envelope} method that retains the sharp FDM cutoff while constraining the high-$k$ plateau depth relevant to mixed scenarios (see Fig.~\ref{fig:Panel_FDM}, right panel). It combines two independently-derived bounds that share the same inference pipeline, data, and likelihood. At low $k$, we adopt the $f_\varphi = 1$ FDM bound from COZMIC~I~\cite{nadler2024cozmicicosmologicalzoomin}, capturing the steep cutoff at MW satellite scales; at high $k$, where COZMIC~I does not constrain the plateau depth relevant to mixed scenarios, we transition to the $\delta = 0.4$ floor from COZMIC~II, the largest $\delta$ ruled out by MW satellite observations.

Although this composite shape has not itself been simulated and data-constrained end-to-end, its construction is well motivated by two recent zoom-in analyses~\cite{Nadler_2025, An_2025}. Specifically, Ref.~\cite{Nadler_2025} showed that subhalo abundances respond locally to features in $P(k)$, and that satellite counts at MW scales are set by a narrow window of wavenumbers near the onset of suppression, at $k \approx 40~\mathrm{Mpc}^{-1}$, corresponding to the minimum halo mass derived from satellites, $\approx 10^8~M_\odot$~\cite{Nadler_2020, An_2025}. Meanwhile, Ref.~\cite{An_2025} showed this onset to be weakly dependent on $\delta$ along its 95\% confidence exclusion boundary for $\delta \leq 0.4$ (see their Fig.~8). Together, these findings motivate using the $\delta = 0.4$ floor as a conservative limit on allowable high-$k$ amplitude features in $P(k)$, independent of the cutoff shape. We thus apply criterion Eq.~(\ref{envelope_criterion}) with $T^2_{\mathrm{env}}$ set to the \textit{composite envelope}, anticipating results of future FDM-tailored simulations.

We evaluate Eq.~(\ref{envelope_criterion}) using candidate beyond-CDM $T^2(k)$s computed with versions of the Boltzmann solver CLASS~\cite{Blas_2011} appropriate to each scenario. For FDM, we use AxiCLASS from Refs.~\cite{Poulin:2018dzj,Smith:2019ihp}.\footnote{\href{https://github.com/PoulinV/AxiCLASS}{https://github.com/PoulinV/AxiCLASS}} For $\gamma$-DM and $p$-DM, we use a modified version of CLASS from Ref.~\cite{Becker_2021}, which accounts for DM sound speed.\footnote{\href{https://github.com/lesgourg/class_public}{CLASS v. 3.2.0.}}  For the $\nu$-DM case, we modify the version used in Ref.~\cite{Mosbech_2021}\footnote{\href{https://github.com/MarkMos/CLASS_nu-DM}{https://github.com/MarkMos/CLASS\_nu-DM}} by including DM sound speed and temperature evolution (see Ref.~\cite{Crumrine_2025}). For IDM we consider DM masses $m_\chi \in [10~\mathrm{keV},\,100~\mathrm{GeV}]$.\footnote{Note that, for $\nu$-DM, we restrict to $m_\chi \geq 100~\mathrm{keV}$, as the $10~\mathrm{keV}$ power spectra exhibit a pronounced rise at $k \gtrsim 100~h~\mathrm{Mpc}^{-1}$. As this may be a numerical artifact, we defer investigation of this regime to future work.}

\section{Results}
\label{sec:results}
Section~\ref{sec:fdm_bounds} reports 95\% confidence lower limits on the FDM scalar field mass, $m_\varphi$, for both \textit{envelope} and \textit{composite envelope} methods, for fractions down to $f_\varphi = 0.5$ (see Table~\ref{tab:fdm}). Section~\ref{sec:idm_bounds} presents 95\% confidence upper limits on IDM cross sections for $\gamma$-DM, $\nu$-DM, and $p$-DM, likewise down to $f_\chi = 0.5$ and across a range of DM masses (see Tables~\ref{tab:photon_dm}--\ref{tab:baryon_dm}). We find that FDM mass bounds relax by a factor of $\sim$1.5, and IDM cross-section bounds by a factor of $\sim$1.9--5.6 (depending on $m_\chi$ and scattering target), from the $100\%$ down to the $50\%$ cases. Note that this $50\%$ floor falls out of our method: at lower fractions, the unsuppressed CDM contribution to $T^2(k)$ prevents any model from satisfying the envelope criterion of Eq.~(\ref{envelope_criterion}).

\begin{figure*}[htbp]\label{fdm}
    \centering
    \includegraphics[width=\textwidth]{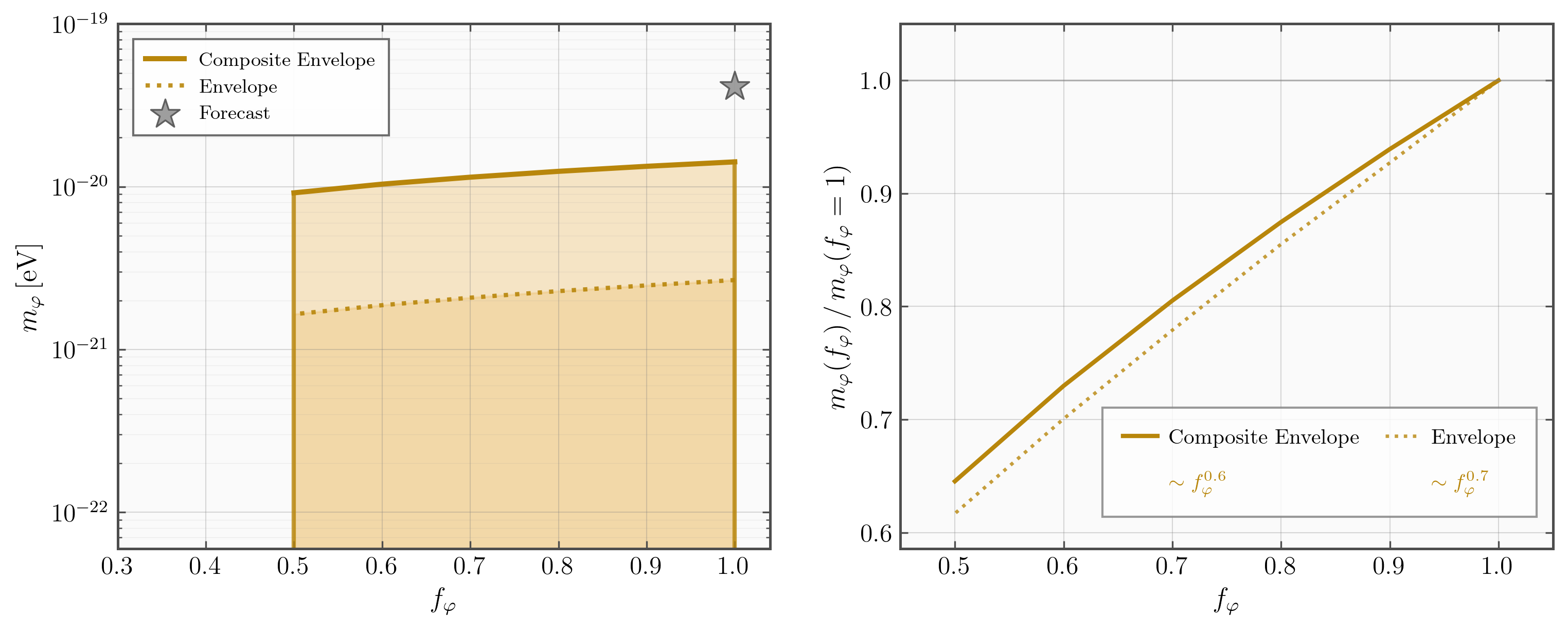}
    \caption{Constraints on mixed FDM. \textit{Left:} Lower limits on the scalar field mass $m_\varphi$ as a function of the FDM fraction $f_\varphi$, from both \textit{envelope} (dotted) and \textit{composite envelope} (solid) methods; the latter represents a more nuanced bound, as explained in the text. Shaded regions are excluded at 95\% confidence by the MW satellite population. The star shows projected sensitivity for the $f_\varphi = 1$ case from the idealized future-satellite-survey forecast of Ref.~\cite{nadler2024forecasts} (Scenario A), obtained by mapping their 95\% confidence WDM limit ($m_{\mathrm{WDM}} > 8.8$~keV) onto FDM parameter space via half-mode matching~\cite{Crumrine_2025}. The forecast assumes complete detection of MW satellites within the virial radius down to approximate LSST sensitivity thresholds, $M_V < 0$~mag and $\mu_V < 32$~mag~arcsec$^{-2}$. \textit{Right:} Scaling of the FDM mass bound with fractional contribution: $m_\varphi(f_\varphi)/m_\varphi(f_\varphi=1)$ versus $f_\varphi$, with fitted power-law $\propto f_\varphi^\beta$.}
    \label{fig:fdm}
\end{figure*}

We additionally report limits on the dimensionless interaction parameters $u_{\chi\gamma}$, $u_{\chi\nu}$, and $u_{\chi p}$ (related to $\sigma_{\chi j}$ via Eq.~(\ref{u})). Furthermore, tabulated in Table~\ref{tab:forecasts} and discussed in Section~\ref{sec:forecasts} are bounds from idealized future-satellite-survey forecasts in the $100\%$ case; see Figures~\ref{fig:fdm}--\ref{fig:baryon} for forecast overlays.

\subsection{FDM Bounds}\label{sec:fdm_bounds}
For the $100\%$ scenario ($f_\varphi = 1$), the \textit{composite envelope} method produces $m_\varphi > 1.41 \times 10^{-20}$~eV, overlaying Ref.~\cite{nadler2024cozmicicosmologicalzoomin}'s bound by construction. The \textit{envelope} method, overly conservative for FDM (see Sec.~\ref{sec:approach}), yields $m_\varphi > 2.65 \times 10^{-21}$~eV, a factor of $\sim$5.3 weaker than its composite counterpart. Bounds from both methods are shown in Fig.~\ref{fig:fdm} for all explored $f_\varphi$, with shaded regions excluded at 95\% confidence. Masses at or below these thresholds place the cutoff in linear power at or below MW satellite wavenumbers, suppressing the formation of observed satellites. Larger masses, corresponding to larger characteristic suppression wavenumbers, remain allowed, as current data do not yet constrain power at these scales.

Compared to other probes, the \textit{composite envelope} bound is $\sim$$40\%$ weaker than the Lyman-$\alpha$ constraint from Ref.~\cite{Rogers_2021}, $\sim20\times$ weaker than the stellar-heating bound from Ref.~\cite{dalal2022fuzzyexcludingfdmsizes}, and $\sim$6$\times$ stronger than the Leo~II kinematic bound from Ref.~\cite{Zimmermann_2025}. The mixed FDM literature, however, focuses on lower fractions and masses, typically $f_\varphi \lesssim 20\%$ for $m_\varphi \sim 10^{-33}$--$10^{-23}\,\mathrm{eV}$~\cite{Hlo_ek_2018,Lagu__2022,Rogers_2023,winch2025highredshiftsmallscaletestsultralight,lazare2025constraintsfuzzydarkmatter}, leaving the high-fraction regime probed here largely unexplored.

For both methods, reducing $f_\varphi$ relaxes the mass bound, with the scaling well described by $m_\varphi / m_\varphi(f_\varphi=1) \propto f_\varphi^{\beta}$, where $\beta \approx 0.6$--$0.7$ depending on method (see Fig.~\ref{fig:fdm}, right panel). Physically, FDM models along the boundary coincide at the same MW-satellite scales in $P(k)$ (Fig.~\ref{fig:Panel_FDM}): from Eq.~\ref{Delta_Tot_FDM}, $T_\mathrm{dm}(k) = (1-f_\varphi) + f_\varphi\, T_{\rm FDM}(k, m_\varphi)$, so a smaller $f_\varphi$ shifts the cutoff to larger scales (smaller $m_\varphi$) to remain on the boundary. The sublinear exponent reflects the mismatch between how $f_\varphi$ enters, as a linear weight, and how $m_\varphi$ enters, through $k_J \propto m_\varphi^{1/2}$. While these scalings with FDM fraction have not, to our knowledge, been characterized in the literature, the qualitative degeneracy direction is consistent with current mixed-FDM analyses at lower fractions~\cite{Rogers_2023,winch2025highredshiftsmallscaletestsultralight,lazare2025constraintsfuzzydarkmatter}.

\begin{figure*}[htbp]\label{photon}
    \centering
    \includegraphics[width=\textwidth]{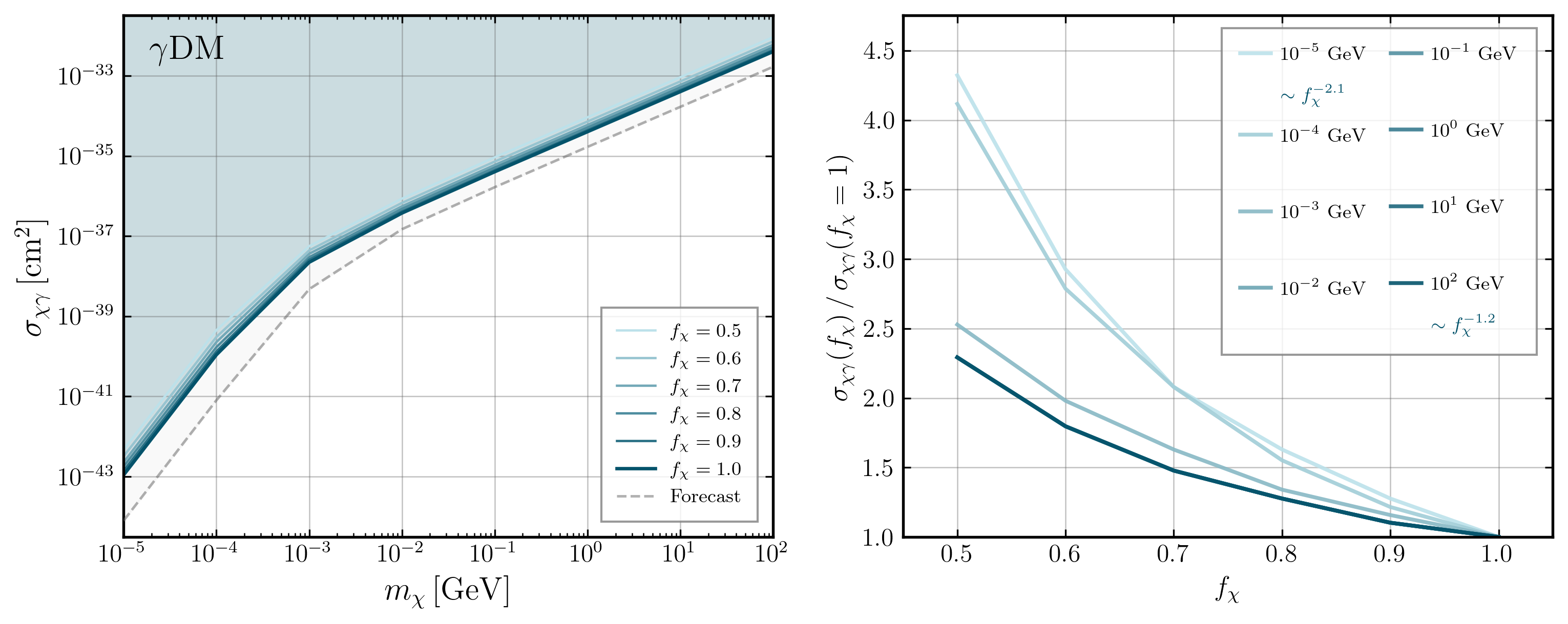}
    \caption{Constraints on mixed $\gamma$-DM+CDM. \textit{Left:} Upper limits on $\sigma_{\chi\gamma}$ as a function of DM mass $m_\chi$ for non-cold fractions $f_\chi = 0.5$--$1.0$. Shaded regions are excluded at 95\% confidence. The dashed line shows projected sensitivity from the idealized future-satellite-survey forecast of Ref.~\cite{nadler2024forecasts} (Scenario A), obtained as in the FDM case, see Fig.~\ref{fig:fdm} caption.  \textit{Right:} Scaling of the cross-section bound with fractional contribution: $\sigma_{\chi\gamma}(f_\chi)/\sigma_{\chi\gamma}(f_\chi=1)$ versus $f_\chi$ for each $m_\chi$, with fitted power-law scalings $\propto f_\chi^{-\alpha}$ indicated for both minimum and maximum tested DM mass.}
    \label{fig:photon}
\end{figure*}
\begin{figure*}[htbp]\label{fig:neutrino}
    \centering
    \includegraphics[width=\textwidth]{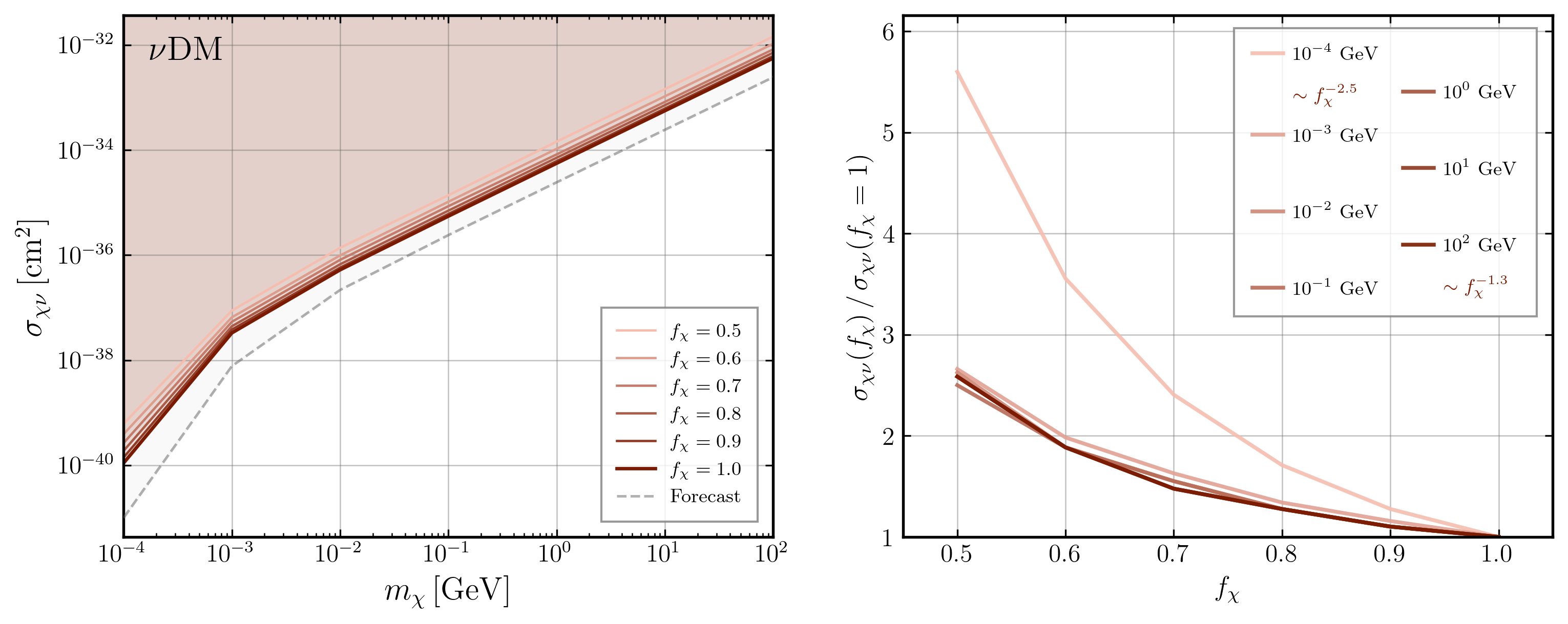}
    \caption{Constraints on mixed $\nu$-DM+CDM. \textit{Left:} Upper limits on $\sigma_{\chi\nu}$ as a function of DM mass $m_\chi$ for non-cold fractions $f_\chi = 0.5$--$1.0$. Shaded regions are excluded at 95\% confidence. The dashed line shows projected sensitivity from the idealized future-satellite-survey forecast of Ref.~\cite{nadler2024forecasts} (Scenario A), obtained as in the FDM case, see Fig.~\ref{fig:fdm} caption. \textit{Right:} Scaling of the cross-section bound with fractional contribution: $\sigma_{\chi\nu}(f_\chi)/\sigma_{\chi\nu}(f_\chi=1)$ versus $f_\chi$ for each $m_\chi$, with fitted power-law scalings $\propto f_\chi^{-\alpha}$ indicated for both minimum and maximum tested DM mass.}
    \label{fig:neutrino}

\end{figure*}
\begin{figure*}[htbp]\label{fig:baryon}
    \centering
    \includegraphics[width=\textwidth]{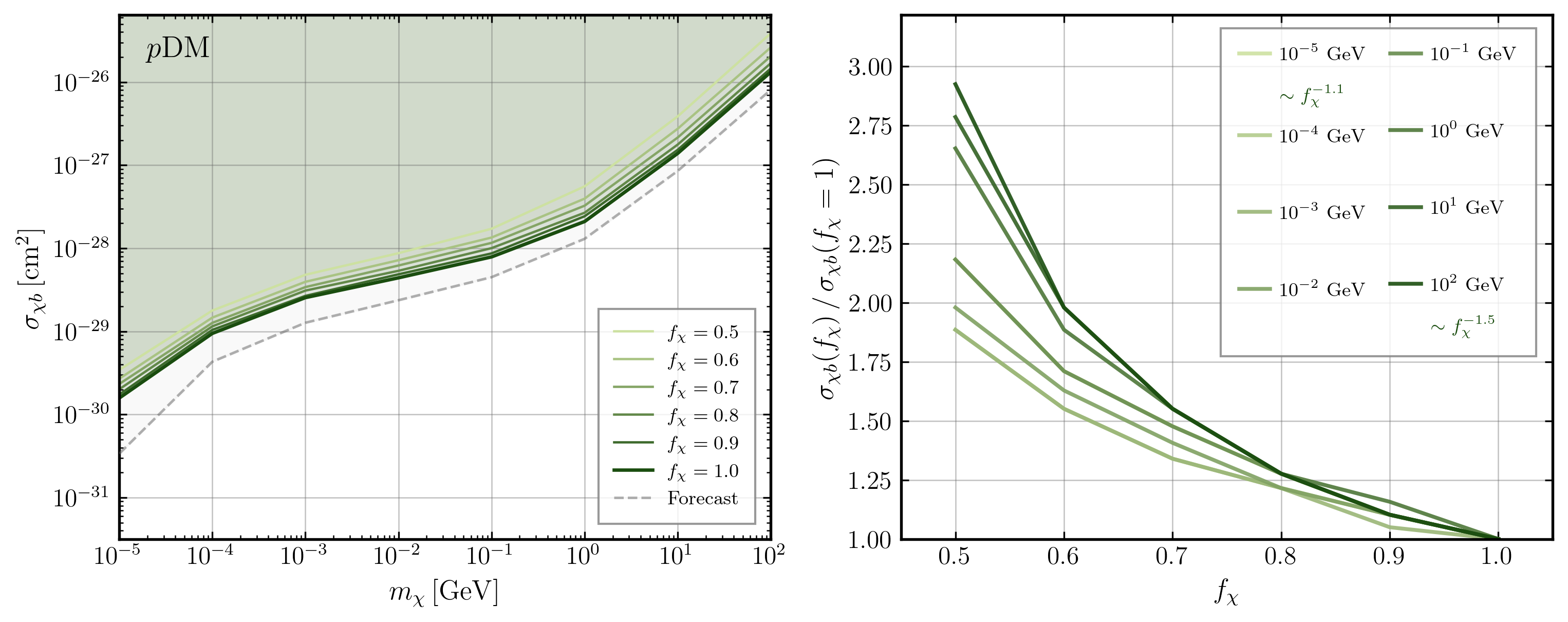}
    \caption{Constraints on mixed $p$-DM+CDM. \textit{Left:} Upper limits on $\sigma_{\chi p}$ as a function of DM mass $m_\chi$ for non-cold fractions $f_\chi = 0.5$--$1.0$. Shaded regions are excluded at 95\% confidence. The dashed line shows projected sensitivity from the idealized future-satellite-survey forecast of Ref.~\cite{nadler2024forecasts} (Scenario A), obtained as in the FDM case, see Fig.~\ref{fig:fdm} caption. \textit{Right:} Scaling of the cross-section bound with fractional contribution: $\sigma_{\chi p}(f_\chi)/\sigma_{\chi p}(f_\chi=1)$ versus $f_\chi$ for each $m_\chi$, with fitted power-law scalings $\propto f_\chi^{-\alpha}$ indicated for both minimum and maximum tested DM mass.}
    \label{fig:baryon}
\end{figure*}

\subsection{IDM Bounds}\label{sec:idm_bounds}

For the $100\%$ scenario at $m_\chi=1~\mathrm{GeV}$, our envelope method yields
$\sigma_{\chi\gamma} < 4.05 \times 10^{-35}~\mathrm{cm}^2$,
$\sigma_{\chi\nu} < 5.59 \times 10^{-35}~\mathrm{cm}^2$, and
$\sigma_{\chi p} < 2.10 \times 10^{-28}~\mathrm{cm}^2$. Bounds for all fractions are shown in Figs.~\ref{fig:photon}--\ref{fig:baryon}, with shaded regions excluded at 95\% confidence. For $p$-DM, our $100\%$ bound improves upon the MW satellite bound from Ref.~\cite{Maamari_2021} by $\sim1$ order of magnitude and the galaxy cluster thermodynamics bound from Ref.~\cite{stuart2025constraintsdarkmatterbaryoninteraction} by $\sim$1--2 orders of magnitude depending on DM mass, and is $\sim$2--3$\times$ weaker than Lyman-$\alpha$ constraints from Ref.~\cite{Rogers_2022} for $m_\chi \gtrsim 100$~keV. Our $100\%$ $\gamma$-DM and $\nu$-DM bounds are consistent with the latest MW bounds from Ref.~\cite{Crumrine_2025}, which improve upon existing CMB, BAO, Lyman-$\alpha$, and weak lensing constraints by $\gtrsim$1--5 orders of magnitude depending on $m_\chi$~\cite{Becker_2021,Hooper_2022,Escudero_2018,Akita_2023,Zu_2026}, with the 5-order improvement driven by DM sound speed effects at low-$m_\chi$. Mixed IDM remains comparatively underexplored; however, several analyses complement ours by pushing to lower fractions: for $p$-DM, from galaxy cluster thermodynamics~\cite{stuart2025constraintsdarkmatterbaryoninteraction}, from Lyman-$\alpha$ and CMB data at $m_\chi = 1$~GeV~\cite{Hooper_2022_Likelihood}, and from CMB data~\cite{straight2026cmbconstraintsdarkmatterproton}; for $\gamma$-DM, from CMB data~\cite{Stadler_2020}. Of these, only the first two reach our fraction range, where our bounds are $\sim$1--2 orders of magnitude and a factor of $\gtrsim$2 tighter, respectively.

We find that decreasing $f_\chi$ permits larger cross sections, following from the density-weighted structure of $\delta_\mathrm{dm}$ (Eq.~(\ref{Delta_Tot_IDM})). We fit power-law scalings $\sigma_{\chi j}/ \sigma_{\chi j}(f_\chi=1) \propto f_\chi^{-\alpha}$ (Figs.~\ref{fig:photon}--\ref{fig:baryon}, right panels), finding the steepest scalings at low $m_\chi$ for radiation scattering and at high $m_\chi$ for proton scattering---suggesting that the regime of greatest sensitivity to $f_\chi$ depends on the scattering target. Unlike FDM, IDM models along the boundary do not always coincide (Fig.~\ref{fig:Panel_9}). The IDM exponents therefore reflect not only how $f_\chi$ enters the density-weighted total perturbation, but also how each model's morphology interacts with the benchmark envelope's shape across MW satellite scales. Nonetheless, the qualitative degeneracy direction we observe is consistent with independent findings from other probes: $\nu$-DM constraints from weak lensing combined with CMB data~\cite{Zu_2026}, and $p$-DM constraints from both galaxy cluster thermodynamics~\cite{stuart2025constraintsdarkmatterbaryoninteraction} and CMB and Lyman-$\alpha$ data~\cite{Hooper_2022_Likelihood}.

\subsection{Future-Satellite-Survey Forecasts}\label{sec:forecasts}
To assess the scope for future improvement on the bounds reported in Secs.~\ref{sec:fdm_bounds} and~\ref{sec:idm_bounds}, we project Ref.~\cite{nadler2024forecasts}'s idealized future-satellite-survey forecasts onto our beyond-CDM parameter spaces. This yields the $100\%$-case limits in Table~\ref{tab:forecasts} (Appendix~\ref{bounds}), appearing also as the star in Fig.~\ref{fig:fdm} and the dashed curves in Figs.~\ref{fig:photon}--\ref{fig:baryon}. These projected limits were obtained by half-mode matching the $95\%$ confidence WDM forecast of Ref.~\cite{nadler2024forecasts} ($m_{\mathrm{WDM}} > 8.8$~keV) to our beyond-CDM transfer functions.\footnote{Half-mode matching equates the half-mode wavenumber $k_{\mathrm{hm}}$ of each candidate FDM or IDM model with that of the forecasted WDM model, isolating the shared scale of linear power suppression; see Ref.~\cite{Crumrine_2025}.} They therefore inherit Ref.~\cite{nadler2024forecasts}'s assumptions of a ``weak'' galaxy formation cutoff and complete detection of MW satellites within the virial radius down to approximate LSST sensitivity thresholds.\footnote{Ref.~\cite{nadler2024forecasts}'s ``weak'' galaxy formation cutoff assumes $M_{50} = 3 \times 10^7\,M_\odot$, where $M_{50}$ is the peak virial mass at which 50\% of halos host detectable galaxies. Their adopted LSST-like sensitivity thresholds are $M_V < 0$~mag and $\mu_V < 32$~mag~arcsec$^{-2}$, where $M_V$ is the absolute $V$-band magnitude and $\mu_V$ the effective surface brightness within the half-light radius.}

For FDM, the forecast tightens our $100\%$-case \textit{envelope} bound by a factor of $\sim$15.5 and the \textit{composite envelope} bound by a factor of $\sim$2.9. For IDM, the forecast tightens our $100\%$-case bounds by a factor of $\sim$1.6--2.4 at high-$m_\chi$ and $\sim$4.7--13.8 at low-$m_\chi$, depending on the scattering species. The disproportionate low-$m_\chi$ improvement reflects the explicit $k$ dependence of the sound speed term ($c_\chi^2 k^2 \delta_\chi$, Eqs.~\ref{boltzmann_baryon_DM_final} and~\ref{boltzmann_rad}): a higher $k_{\mathrm{hm}}$ enhances pressure support, extending the pressure-dominated regime to higher $m_\chi$~\cite{Crumrine_2025}.

\section{Discussion and Conclusions}
\label{sec:conclusion}

Interactions between DM and SM particles, as well as ultralight scalar field dynamics, suppress linear matter perturbations in the early Universe and thereby reduce the abundance of present-day dwarf galaxies. We explore mixed DM scenarios in which only a fraction of DM exhibits non-CDM behavior, while the remainder behaves as CDM. Using the observed MW satellite population, we set new leading constraints on four such cases: fuzzy DM and DM that elastically scatters with photons, neutrinos, or protons. We do this by mapping power-spectrum limits from the COZMIC suite of beyond-CDM zoom-in simulations~\cite{nadler2024cozmicicosmologicalzoomin, An_2025} onto each parameter space, a procedure that bypasses the need for additional dedicated simulations of each scenario. 

This work probes non-CDM fractions down to $50\%$ for FDM and all three IDM interaction types, a region of parameter space that remains sparsely explored by both MW satellite analyses and other cosmological probes over the DM masses considered. We find that FDM mass bounds relax by a factor of $\sim$1.5, and IDM cross-section bounds by a factor of $\sim$2--6, as the non-CDM fraction drops from $100\%$ to $50\%$. This follows from the density-weighted structure of the total DM perturbation (Eqs.~\ref{Delta_Tot_FDM} and~\ref{Delta_Tot_IDM}): a smaller non-CDM contribution requires stronger per-particle effects to produce the same suppression signature in the MW satellite population. Empirical power-law scalings (Figs.~\ref{fig:fdm}--\ref{fig:baryon}, right panels) capture this fraction dependence through scenario-specific scaling exponents. Method dependence is likely a general feature of such scalings (see Secs.~\ref{sec:fdm_bounds} and~\ref{sec:idm_bounds}), making our fits a first-order characterization of the degeneracy with IDM or FDM fraction. A robust treatment awaits self-consistent simulations, which will also verify whether late-time effects absent from our linear-theory mapping, such as residual IDM scattering and FDM wave dynamics, remain negligible for subhalo abundances.

Upcoming facilities such as the Vera C. Rubin Observatory and the Nancy Grace Roman Space Telescope promise to reveal faint dwarf galaxies throughout the Local Volume. Access to smaller scales and a larger satellite census will tighten bounds on interaction strengths and DM mass and reduce uncertainties on the faint-end satellite luminosity function, enabling probes of lower beyond-CDM fractions~\cite{nadler2024forecasts, banerjee2022snowmass2021cosmicfrontierwhite, boddy2022astrophysicalcosmologicalprobesdark, mao2022snowmass2021veracrubin, bechtol2023snowmass2021cosmicfrontierwhite, LSSTDarkMatterGroup:2019mwo, chakrabarti2022snowmass2021cosmicfrontierwhite}. By translating idealized future-satellite-survey forecasts of Ref.~\cite{nadler2024forecasts} onto our beyond-CDM parameter spaces (see Section~\ref{sec:forecasts}), we project improvement on $100\%$-case bounds by a factor of $\sim$1.6--14 for IDM and $\sim$3 for FDM (overlays in Figs.~\ref{fig:fdm}--\ref{fig:baryon}), assuming complete satellite detection within the MW virial radius, down to approximate LSST sensitivity thresholds. More realistic projections of bound improvement must fold in LSST's sky footprint and faint-end detection efficiency, while requiring object and/or population-level methods to disentangle dwarf galaxies from star cluster contaminants.

Independent probes beyond MW satellites can offer complementary tests of mixed FDM and IDM scenarios. One such avenue could target the distinctive growth signature of some mixed DM models, in which the high-$k$ tail suppression deepens relative to $\Lambda$CDM over cosmic time (see Appendix~\ref{morphology} for a discussion of this physics). Ratios of $P(k)$ at different redshifts could isolate this growth deficit on larger scales, yielding constraints from galaxy surveys or line intensity mapping analyses~\cite{Bernal_2022}. Combined with advances in CMB measurements at small angular scales, strong gravitational lensing, Lyman-$\alpha$ forest analyses, and galaxy clustering, a multi-probe approach will be essential to constrain---or discover---mixed DM physics~\cite{SimonsObservatory:2018koc,CMB-S4:2022ght, Keeley_2023, gluscevic2019cosmological, boddy2022astrophysicalcosmologicalprobesdark}.

\section*{Acknowledgments}
W.C. acknowledges Adrienne Erickcek and Keir Rogers for helpful discussions about mixed DM during the Fall 2025 Spec-S5 workshop.  This material is based upon work supported by the National
Science Foundation under grant No. 2509561 (E.O.N. and
A.B.) and No. 2407380 (V.G.). Any opinions, findings, and
conclusions or recommendations expressed in this material are
those of the author(s) and do not necessarily reflect the views
of the NSF. V.G. also acknowledges the support from the National Science Foundation (NSF) CAREER Grant No. PHY-2239205, from the Research Corporation for Science Advancement under the Cottrell Scholar Program, from the IBM Einstein Fellowship at the Institute for Advanced Study, and from the Grant 63667 from the John Templeton Foundation. The opinions expressed in this publication are those of the author(s) and do not necessarily reflect the views of the John Templeton Foundation. This research was supported in part by Grant NSF PHY-2309135 to the Kavli Institute for Theoretical Physics (KITP).

\bibliography{main}
\clearpage

\appendix

\section{High $k$ Morphology and Damping Regimes}\label{morphology}

\begin{figure*}[!t]
    \centering
    \includegraphics[width=\textwidth]{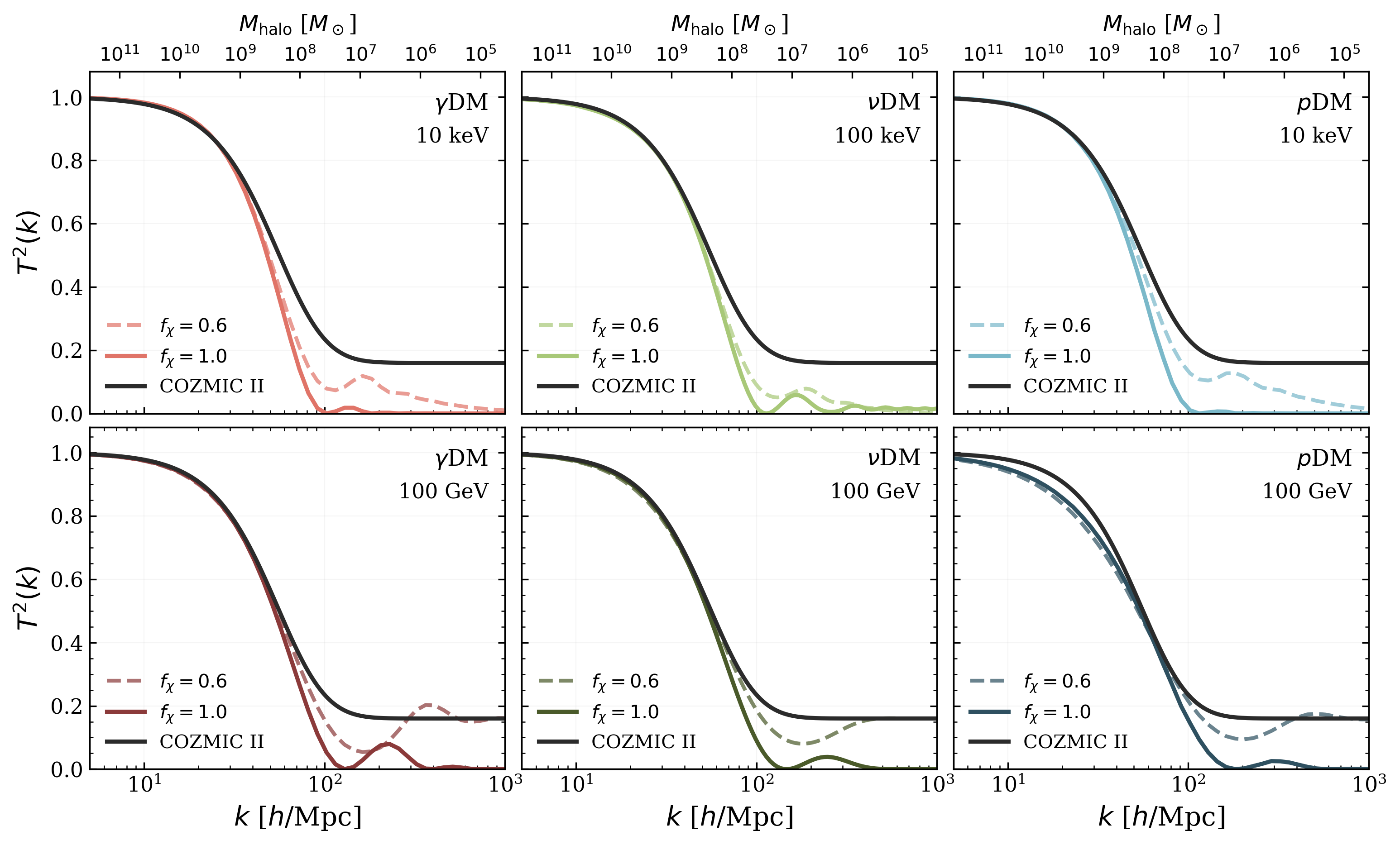}
    \caption{High-$k$ asymptotic behavior of mixed IDM transfer functions. Top row: low-mass (pressure-dominated) regime; bottom row: high-mass (drag-dominated) regime. Columns show $\gamma$-DM (left), $\nu$-DM (center), and $p$-DM (right). Solid and dashed curves correspond to $f_\chi = 1.0$ and $0.6$, respectively, with the reference envelope in black solid. At high $m_\chi$, transfer functions plateau to $(1-f_\chi)^2$; at low $m_\chi$, power continues to decline at high $k$.}
    \label{fig:Panel_6}
\end{figure*}

Our analysis reveals distinct asymptotic signatures in transfer function morphology that reflect underlying damping physics. Fig.~\ref{fig:Panel_6} examines the small-scale behavior of mixed transfer functions at representative high and low DM mass, for two values of $f_\chi$. At high $m_\chi$, mixed $T^2(k)$'s plateau to $(1 - f_\chi)^2$, while at low $m_\chi$, as well as in FDM (see Fig.~\ref{fig:Panel_FDM}) and in WDM (see Ref.~\cite{An_2025}), power continues to fall toward complete suppression at high $k$. 

This behavior arises from a given scenario's dominant damping mechanism. In the drag-dominated regime of collisional damping (IDM, high $m_\chi$), perturbation erasure halts at all $k$ simultaneously upon decoupling, with no further suppression thereafter; IDM then evolves as a pressureless fluid alongside CDM, locking $T^2(k)$ at $(1-f_\chi)^2$. In pressure-dominated scenarios (low-$m_\chi$ IDM, FDM and WDM), the relevant suppression terms in the Boltzmann equations carry explicit $k$-dependence—quantum pressure for FDM, thermal pressure for WDM and low-$m_\chi$ IDM—and this pressure support, though confined to progressively smaller scales as the effective Jeans scale shrinks with time, persists post-decoupling. The continued suppression of small-scale clustering further weakens metric perturbations, slowing even CDM growth and pushing the transfer function below $(1-f_\chi)^2$. 

We have verified this cumulative growth deficit numerically across redshift: the high-$k$ tail suppression relative to $\Lambda$CDM deepens over cosmic time, suggesting an avenue for future work probing a redshift-dependent imprint on $P(k)$.

\clearpage

\section{Tabulated Bounds and Forecasts}\label{bounds}

\vspace{1.4em}
\begin{table}[htbp]
\centering
\renewcommand{\arraystretch}{1.5}
\begin{tabular}{crr}
\toprule
\multicolumn{3}{c}{\textbf{Fuzzy Dark Matter (FDM)}} \\
\midrule
& \multicolumn{1}{l}{\textbf{Envelope}} & \textbf{Composite Envelope} \\
$f_\varphi$ & $m_\varphi$ [$10^{-22}$ eV] & $m_\varphi$ [$10^{-22}$ eV] \\
\midrule
1.00 & 26.5 & 141 \\
0.90 & 24.6 & 132 \\
0.80 & 22.7 & 123 \\
0.70 & 20.6 & 113 \\
0.60 & 18.6 & 103 \\
0.50 & 16.3 & 91.0 \\
\bottomrule
\end{tabular}
\caption{Lower limits on the FDM scalar field mass $m_\varphi$ at 95\% confidence for interaction fractions $f_\varphi = 0.5$--$1.0$, derived using both \textit{envelope} and \textit{composite envelope} methods.}
\label{tab:fdm}
\end{table}

\begin{table}[htbp]\label{photon_table}
\centering
\renewcommand{\arraystretch}{1.2}
\setlength{\tabcolsep}{12pt}
\begin{tabular}{cccc}
\toprule
\multicolumn{4}{c}{\textbf{Photon Scattering ($\gamma$-DM)}} \\
\midrule
$m_\chi$ [GeV] & $f_\chi$ & $u_{\chi\gamma}$ & $\sigma_{\chi\gamma}$ [cm$^2$] \\
\midrule
$10^{2}$ & 1.0 & $6.09\times10^{-9}$ & $4.05\times10^{-33}$ \\
$10^{2}$ & 0.9 & $6.71\times10^{-9}$ & $4.46\times10^{-33}$ \\
$10^{2}$ & 0.8 & $7.77\times10^{-9}$ & $5.17\times10^{-33}$ \\
$10^{2}$ & 0.7 & $9.00\times10^{-9}$ & $5.98\times10^{-33}$ \\
$10^{2}$ & 0.6 & $1.09\times10^{-8}$ & $7.28\times10^{-33}$ \\
$10^{2}$ & 0.5 & $1.40\times10^{-8}$ & $9.28\times10^{-33}$ \\
\midrule
$10^{1}$ & 1.0 & $6.09\times10^{-9}$ & $4.05\times10^{-34}$ \\
$10^{1}$ & 0.9 & $6.71\times10^{-9}$ & $4.46\times10^{-34}$ \\
$10^{1}$ & 0.8 & $7.77\times10^{-9}$ & $5.17\times10^{-34}$ \\
$10^{1}$ & 0.7 & $9.00\times10^{-9}$ & $5.98\times10^{-34}$ \\
$10^{1}$ & 0.6 & $1.09\times10^{-8}$ & $7.28\times10^{-34}$ \\
$10^{1}$ & 0.5 & $1.40\times10^{-8}$ & $9.28\times10^{-34}$ \\
\midrule
$10^{0}$ & 1.0 & $6.09\times10^{-9}$ & $4.05\times10^{-35}$ \\
$10^{0}$ & 0.9 & $6.71\times10^{-9}$ & $4.46\times10^{-35}$ \\
$10^{0}$ & 0.8 & $7.77\times10^{-9}$ & $5.17\times10^{-35}$ \\
$10^{0}$ & 0.7 & $9.00\times10^{-9}$ & $5.98\times10^{-35}$ \\
$10^{0}$ & 0.6 & $1.09\times10^{-8}$ & $7.28\times10^{-35}$ \\
$10^{0}$ & 0.5 & $1.40\times10^{-8}$ & $9.28\times10^{-35}$ \\
\midrule
$10^{-1}$ & 1.0 & $6.09\times10^{-9}$ & $4.05\times10^{-36}$ \\
$10^{-1}$ & 0.9 & $6.71\times10^{-9}$ & $4.46\times10^{-36}$ \\
$10^{-1}$ & 0.8 & $7.77\times10^{-9}$ & $5.17\times10^{-36}$ \\
$10^{-1}$ & 0.7 & $9.00\times10^{-9}$ & $5.98\times10^{-36}$ \\
$10^{-1}$ & 0.6 & $1.09\times10^{-8}$ & $7.28\times10^{-36}$ \\
$10^{-1}$ & 0.5 & $1.40\times10^{-8}$ & $9.28\times10^{-36}$ \\
\midrule
$10^{-2}$ & 1.0 & $5.67\times10^{-9}$ & $3.77\times10^{-37}$ \\
$10^{-2}$ & 0.9 & $6.25\times10^{-9}$ & $4.16\times10^{-37}$ \\
$10^{-2}$ & 0.8 & $7.24\times10^{-9}$ & $4.81\times10^{-37}$ \\
$10^{-2}$ & 0.7 & $8.38\times10^{-9}$ & $5.57\times10^{-37}$ \\
$10^{-2}$ & 0.6 & $1.02\times10^{-8}$ & $6.77\times10^{-37}$ \\
$10^{-2}$ & 0.5 & $1.30\times10^{-8}$ & $8.65\times10^{-37}$ \\
\midrule
$10^{-3}$ & 1.0 & $3.35\times10^{-9}$ & $2.23\times10^{-38}$ \\
$10^{-3}$ & 0.9 & $3.88\times10^{-9}$ & $2.58\times10^{-38}$ \\
$10^{-3}$ & 0.8 & $4.49\times10^{-9}$ & $2.99\times10^{-38}$ \\
$10^{-3}$ & 0.7 & $5.46\times10^{-9}$ & $3.63\times10^{-38}$ \\
$10^{-3}$ & 0.6 & $6.63\times10^{-9}$ & $4.41\times10^{-38}$ \\
$10^{-3}$ & 0.5 & $8.47\times10^{-9}$ & $5.63\times10^{-38}$ \\
\midrule
$10^{-4}$ & 1.0 & $1.64\times10^{-10}$ & $1.09\times10^{-40}$ \\
$10^{-4}$ & 0.9 & $2.00\times10^{-10}$ & $1.33\times10^{-40}$ \\
$10^{-4}$ & 0.8 & $2.55\times10^{-10}$ & $1.70\times10^{-40}$ \\
$10^{-4}$ & 0.7 & $3.42\times10^{-10}$ & $2.27\times10^{-40}$ \\
$10^{-4}$ & 0.6 & $4.58\times10^{-10}$ & $3.04\times10^{-40}$ \\
$10^{-4}$ & 0.5 & $6.76\times10^{-10}$ & $4.50\times10^{-40}$ \\
\midrule
$10^{-5}$ & 1.0 & $1.64\times10^{-12}$ & $1.09\times10^{-43}$ \\
$10^{-5}$ & 0.9 & $2.10\times10^{-12}$ & $1.39\times10^{-43}$ \\
$10^{-5}$ & 0.8 & $2.68\times10^{-12}$ & $1.78\times10^{-43}$ \\
$10^{-5}$ & 0.7 & $3.42\times10^{-12}$ & $2.27\times10^{-43}$ \\
$10^{-5}$ & 0.6 & $4.81\times10^{-12}$ & $3.20\times10^{-43}$ \\
$10^{-5}$ & 0.5 & $7.10\times10^{-12}$ & $4.72\times10^{-43}$ \\
\bottomrule
\end{tabular}
\caption{Upper limits on the $\gamma$-DM interaction cross section $\sigma_{\chi\gamma}$ and dimensionless parameter $u_{\chi\gamma}$ at 95\% confidence, for DM masses $m_\chi = 10^{-5}$--$10^2$~GeV and fractions $f_\chi = 0.5$--$1.0$.}
\label{tab:photon_dm}
\end{table}

\begin{table}[htbp]\label{neutrino_table}
\centering
\renewcommand{\arraystretch}{1.2}
\setlength{\tabcolsep}{12pt}
\begin{tabular}{cccc}
\toprule
\multicolumn{4}{c}{\textbf{Neutrino Scattering ($\nu$-DM)}} \\
\midrule
$m_\chi$ [GeV] & $f_\chi$ & $u_{\chi\nu}$ & $\sigma_{\chi\nu}$ [cm$^2$] \\
\midrule
$10^{2}$ & 1.0 & $8.40\times10^{-9}$ & $5.59\times10^{-33}$ \\
$10^{2}$ & 0.9 & $9.26\times10^{-9}$ & $6.16\times10^{-33}$ \\
$10^{2}$ & 0.8 & $1.07\times10^{-8}$ & $7.13\times10^{-33}$ \\
$10^{2}$ & 0.7 & $1.24\times10^{-8}$ & $8.25\times10^{-33}$ \\
$10^{2}$ & 0.6 & $1.58\times10^{-8}$ & $1.05\times10^{-32}$ \\
$10^{2}$ & 0.5 & $2.17\times10^{-8}$ & $1.44\times10^{-32}$ \\
\midrule
$10^{1}$ & 1.0 & $8.40\times10^{-9}$ & $5.59\times10^{-34}$ \\
$10^{1}$ & 0.9 & $9.26\times10^{-9}$ & $6.16\times10^{-34}$ \\
$10^{1}$ & 0.8 & $1.07\times10^{-8}$ & $7.13\times10^{-34}$ \\
$10^{1}$ & 0.7 & $1.24\times10^{-8}$ & $8.25\times10^{-34}$ \\
$10^{1}$ & 0.6 & $1.58\times10^{-8}$ & $1.05\times10^{-33}$ \\
$10^{1}$ & 0.5 & $2.17\times10^{-8}$ & $1.44\times10^{-33}$ \\
\midrule
$10^{0}$ & 1.0 & $8.40\times10^{-9}$ & $5.59\times10^{-35}$ \\
$10^{0}$ & 0.9 & $9.26\times10^{-9}$ & $6.16\times10^{-35}$ \\
$10^{0}$ & 0.8 & $1.07\times10^{-8}$ & $7.13\times10^{-35}$ \\
$10^{0}$ & 0.7 & $1.24\times10^{-8}$ & $8.25\times10^{-35}$ \\
$10^{0}$ & 0.6 & $1.58\times10^{-8}$ & $1.05\times10^{-34}$ \\
$10^{0}$ & 0.5 & $2.17\times10^{-8}$ & $1.44\times10^{-34}$ \\
\midrule
$10^{-1}$ & 1.0 & $8.27\times10^{-9}$ & $5.50\times10^{-36}$ \\
$10^{-1}$ & 0.9 & $9.12\times10^{-9}$ & $6.06\times10^{-36}$ \\
$10^{-1}$ & 0.8 & $1.06\times10^{-8}$ & $7.02\times10^{-36}$ \\
$10^{-1}$ & 0.7 & $1.28\times10^{-8}$ & $8.53\times10^{-36}$ \\
$10^{-1}$ & 0.6 & $1.56\times10^{-8}$ & $1.04\times10^{-35}$ \\
$10^{-1}$ & 0.5 & $2.07\times10^{-8}$ & $1.38\times10^{-35}$ \\
\midrule
$10^{-2}$ & 1.0 & $7.88\times10^{-9}$ & $5.24\times10^{-37}$ \\
$10^{-2}$ & 0.9 & $8.68\times10^{-9}$ & $5.77\times10^{-37}$ \\
$10^{-2}$ & 0.8 & $1.01\times10^{-8}$ & $6.68\times10^{-37}$ \\
$10^{-2}$ & 0.7 & $1.22\times10^{-8}$ & $8.13\times10^{-37}$ \\
$10^{-2}$ & 0.6 & $1.49\times10^{-8}$ & $9.88\times10^{-37}$ \\
$10^{-2}$ & 0.5 & $2.07\times10^{-8}$ & $1.38\times10^{-36}$ \\
\midrule
$10^{-3}$ & 1.0 & $4.96\times10^{-9}$ & $3.30\times10^{-38}$ \\
$10^{-3}$ & 0.9 & $5.74\times10^{-9}$ & $3.82\times10^{-38}$ \\
$10^{-3}$ & 0.8 & $6.65\times10^{-9}$ & $4.42\times10^{-38}$ \\
$10^{-3}$ & 0.7 & $8.08\times10^{-9}$ & $5.37\times10^{-38}$ \\
$10^{-3}$ & 0.6 & $9.84\times10^{-9}$ & $6.54\times10^{-38}$ \\
$10^{-3}$ & 0.5 & $1.32\times10^{-8}$ & $8.78\times10^{-38}$ \\
\midrule
$10^{-4}$ & 1.0 & $1.65\times10^{-10}$ & $1.10\times10^{-40}$ \\
$10^{-4}$ & 0.9 & $2.11\times10^{-10}$ & $1.40\times10^{-40}$ \\
$10^{-4}$ & 0.8 & $2.83\times10^{-10}$ & $1.88\times10^{-40}$ \\
$10^{-4}$ & 0.7 & $3.98\times10^{-10}$ & $2.65\times10^{-40}$ \\
$10^{-4}$ & 0.6 & $5.88\times10^{-10}$ & $3.91\times10^{-40}$ \\
$10^{-4}$ & 0.5 & $9.26\times10^{-10}$ & $6.16\times10^{-40}$ \\
\bottomrule
\end{tabular}
\caption{Upper limits on the $\nu$-DM interaction cross section $\sigma_{\chi\nu}$ and dimensionless parameter $u_{\chi\nu}$ at 95\% confidence, for DM masses $m_\chi = 10^{-4}$--$10^2$~GeV and fractions $f_\chi = 0.5$--$1.0$.}
\label{tab:neutrino_dm}
\end{table}

\begin{table}[htbp]\label{proton_table}
\centering
\renewcommand{\arraystretch}{1.2}
\setlength{\tabcolsep}{12pt}
\begin{tabular}{cccc}
\toprule
\multicolumn{4}{c}{\textbf{Proton Scattering ($p$-DM)}} \\
\midrule
$m_\chi$ [GeV] & $f_\chi$ & $u_{\chi p}$ & $\sigma_{\chi p}$ [cm$^2$] \\
\midrule
$10^{2}$ & 1.0 & $1.99\times10^{-2}$ & $1.32\times10^{-26}$ \\
$10^{2}$ & 0.9 & $2.19\times10^{-2}$ & $1.46\times10^{-26}$ \\
$10^{2}$ & 0.8 & $2.54\times10^{-2}$ & $1.69\times10^{-26}$ \\
$10^{2}$ & 0.7 & $3.09\times10^{-2}$ & $2.05\times10^{-26}$ \\
$10^{2}$ & 0.6 & $3.94\times10^{-2}$ & $2.62\times10^{-26}$ \\
$10^{2}$ & 0.5 & $5.82\times10^{-2}$ & $3.87\times10^{-26}$ \\
\midrule
$10^{1}$ & 1.0 & $2.09\times10^{-2}$ & $1.39\times10^{-27}$ \\
$10^{1}$ & 0.9 & $2.30\times10^{-2}$ & $1.53\times10^{-27}$ \\
$10^{1}$ & 0.8 & $2.67\times10^{-2}$ & $1.77\times10^{-27}$ \\
$10^{1}$ & 0.7 & $3.24\times10^{-2}$ & $2.16\times10^{-27}$ \\
$10^{1}$ & 0.6 & $4.14\times10^{-2}$ & $2.75\times10^{-27}$ \\
$10^{1}$ & 0.5 & $5.82\times10^{-2}$ & $3.87\times10^{-27}$ \\
\midrule
$10^{0}$ & 1.0 & $3.16\times10^{-2}$ & $2.10\times10^{-28}$ \\
$10^{0}$ & 0.9 & $3.66\times10^{-2}$ & $2.43\times10^{-28}$ \\
$10^{0}$ & 0.8 & $4.03\times10^{-2}$ & $2.68\times10^{-28}$ \\
$10^{0}$ & 0.7 & $4.90\times10^{-2}$ & $3.26\times10^{-28}$ \\
$10^{0}$ & 0.6 & $5.95\times10^{-2}$ & $3.96\times10^{-28}$ \\
$10^{0}$ & 0.5 & $8.38\times10^{-2}$ & $5.57\times10^{-28}$ \\
\midrule
$10^{-1}$ & 1.0 & $1.18\times10^{-1}$ & $7.88\times10^{-29}$ \\
$10^{-1}$ & 0.9 & $1.31\times10^{-1}$ & $8.68\times10^{-29}$ \\
$10^{-1}$ & 0.8 & $1.51\times10^{-1}$ & $1.01\times10^{-28}$ \\
$10^{-1}$ & 0.7 & $1.75\times10^{-1}$ & $1.16\times10^{-28}$ \\
$10^{-1}$ & 0.6 & $2.03\times10^{-1}$ & $1.35\times10^{-28}$ \\
$10^{-1}$ & 0.5 & $2.58\times10^{-1}$ & $1.72\times10^{-28}$ \\
\midrule
$10^{-2}$ & 1.0 & $6.63\times10^{-1}$ & $4.41\times10^{-29}$ \\
$10^{-2}$ & 0.9 & $7.31\times10^{-1}$ & $4.86\times10^{-29}$ \\
$10^{-2}$ & 0.8 & $8.06\times10^{-1}$ & $5.36\times10^{-29}$ \\
$10^{-2}$ & 0.7 & $9.33\times10^{-1}$ & $6.21\times10^{-29}$ \\
$10^{-2}$ & 0.6 & $1.08\times10^{0}$ & $7.18\times10^{-29}$ \\
$10^{-2}$ & 0.5 & $1.31\times10^{0}$ & $8.73\times10^{-29}$ \\
\midrule
$10^{-3}$ & 1.0 & $3.83\times10^{0}$ & $2.55\times10^{-29}$ \\
$10^{-3}$ & 0.9 & $4.02\times10^{0}$ & $2.67\times10^{-29}$ \\
$10^{-3}$ & 0.8 & $4.66\times10^{0}$ & $3.10\times10^{-29}$ \\
$10^{-3}$ & 0.7 & $5.13\times10^{0}$ & $3.41\times10^{-29}$ \\
$10^{-3}$ & 0.6 & $5.94\times10^{0}$ & $3.95\times10^{-29}$ \\
$10^{-3}$ & 0.5 & $7.22\times10^{0}$ & $4.80\times10^{-29}$ \\
\midrule
$10^{-4}$ & 1.0 & $1.42\times10^{1}$ & $9.45\times10^{-30}$ \\
$10^{-4}$ & 0.9 & $1.57\times10^{1}$ & $1.04\times10^{-29}$ \\
$10^{-4}$ & 0.8 & $1.73\times10^{1}$ & $1.15\times10^{-29}$ \\
$10^{-4}$ & 0.7 & $1.90\times10^{1}$ & $1.27\times10^{-29}$ \\
$10^{-4}$ & 0.6 & $2.20\times10^{1}$ & $1.47\times10^{-29}$ \\
$10^{-4}$ & 0.5 & $2.68\times10^{1}$ & $1.78\times10^{-29}$ \\
\midrule
$10^{-5}$ & 1.0 & $2.37\times10^{1}$ & $1.58\times10^{-30}$ \\
$10^{-5}$ & 0.9 & $2.61\times10^{1}$ & $1.74\times10^{-30}$ \\
$10^{-5}$ & 0.8 & $3.02\times10^{1}$ & $2.01\times10^{-30}$ \\
$10^{-5}$ & 0.7 & $3.50\times10^{1}$ & $2.33\times10^{-30}$ \\
$10^{-5}$ & 0.6 & $4.05\times10^{1}$ & $2.69\times10^{-30}$ \\
$10^{-5}$ & 0.5 & $5.17\times10^{1}$ & $3.44\times10^{-30}$ \\
\bottomrule
\end{tabular}
\caption{Upper limits on the $p$-DM interaction cross section $\sigma_{\chi p}$ and dimensionless parameter $u_{\chi p}$ at 95\% confidence, for DM masses $m_\chi = 10^{-5}$--$10^2$~GeV and fractions $f_\chi = 0.5$--$1.0$.}
\label{tab:baryon_dm}
\end{table}

\begin{table}[htbp]
\centering
\renewcommand{\arraystretch}{1.2}
\setlength{\tabcolsep}{10pt}
\begin{tabular}{lccc}
\toprule
\multicolumn{4}{c}{\textbf{Idealized Future Satellite Survey Forecasts}} \\
\midrule[\heavyrulewidth]
\multicolumn{4}{c}{\textbf{\textit{FDM}}} \\
\midrule
\multicolumn{4}{c}{$m_\varphi > 411\times10^{-22}$~eV} \\
\midrule[\heavyrulewidth]
\multicolumn{4}{c}{\textbf{\textit{IDM}}} \\
\midrule
Scenario & $m_\chi$ [GeV] & $u_{\chi j}$ & $\sigma_{\chi j}$ [cm$^2$] \\
\midrule
\multirow{8}{*}{$\gamma$-DM}
& $10^{-5}$ & $1.18\times10^{-13}$ & $7.90\times10^{-45}$ \\
& $10^{-4}$ & $1.20\times10^{-11}$ & $8.00\times10^{-42}$ \\
& $10^{-3}$ & $7.14\times10^{-10}$ & $4.75\times10^{-39}$ \\
& $10^{-2}$ & $2.24\times10^{-9}$  & $1.49\times10^{-37}$ \\
& $10^{-1}$ & $2.51\times10^{-9}$  & $1.67\times10^{-36}$ \\
& $10^{0}$  & $2.54\times10^{-9}$  & $1.69\times10^{-35}$ \\
& $10^{1}$  & $2.54\times10^{-9}$  & $1.69\times10^{-34}$ \\
& $10^{2}$  & $2.54\times10^{-9}$  & $1.69\times10^{-33}$ \\
\midrule
\multirow{7}{*}{$\nu$-DM}
& $10^{-4}$ & $1.50\times10^{-11}$ & $9.98\times10^{-42}$ \\
& $10^{-3}$ & $1.16\times10^{-9}$  & $7.71\times10^{-39}$ \\
& $10^{-2}$ & $3.26\times10^{-9}$  & $2.17\times10^{-37}$ \\
& $10^{-1}$ & $3.61\times10^{-9}$  & $2.40\times10^{-36}$ \\
& $10^{0}$  & $3.65\times10^{-9}$  & $2.43\times10^{-35}$ \\
& $10^{1}$  & $3.65\times10^{-9}$  & $2.43\times10^{-34}$ \\
& $10^{2}$  & $3.65\times10^{-9}$  & $2.43\times10^{-33}$ \\
\midrule
\multirow{8}{*}{$p$-DM}
& $10^{-5}$ & $5.06\times10^{0}$   & $3.37\times10^{-31}$ \\
& $10^{-4}$ & $6.48\times10^{0}$   & $4.31\times10^{-30}$ \\
& $10^{-3}$ & $1.92\times10^{0}$   & $1.27\times10^{-29}$ \\
& $10^{-2}$ & $3.56\times10^{-1}$  & $2.37\times10^{-29}$ \\
& $10^{-1}$ & $6.75\times10^{-2}$  & $4.49\times10^{-29}$ \\
& $10^{0}$  & $1.95\times10^{-2}$  & $1.30\times10^{-28}$ \\
& $10^{1}$  & $1.28\times10^{-2}$  & $8.53\times10^{-28}$ \\
& $10^{2}$  & $1.22\times10^{-2}$  & $8.09\times10^{-27}$ \\
\bottomrule
\end{tabular}
\caption{Projected 95\% confidence lower limit on $m_\varphi$ for FDM at $f_\varphi = 1$, and upper limits on $\sigma_{\chi j}$ and $u_{\chi j}$ for $\gamma$-DM, $\nu$-DM, and $p$-DM at $f_\chi = 1$, obtained by mapping the 95\% confidence WDM limit ($m_\mathrm{WDM} > 8.8$~keV) from the idealized future-satellite-survey forecast of Ref.~\cite{nadler2024forecasts} (Scenario A) onto each scenario's parameter space via half-mode matching~\cite{Crumrine_2025}. The forecast assumes complete detection of MW satellites within the virial radius down to approximate LSST sensitivity thresholds, $M_V < 0$~mag and $\mu_V < 32$~mag~arcsec$^{-2}$.}
\label{tab:forecasts}
\end{table}
\clearpage

\end{document}